\pdfoutput=1
\documentclass[twocolumn,10pt]{article}
\usepackage[a4paper,margin=0.7in]{geometry}
\usepackage[T1]{fontenc}
\usepackage[small,compact]{titlesec}
\titleformat{\section}{\normalfont\large\bfseries}{\thesection}{0.7em}{}
\titleformat{\subsection}{\normalfont\normalsize\bfseries}{\thesubsection}{0.6em}{}
\titleformat{\subsubsection}{\normalfont\normalsize\itshape}{\thesubsubsection}{0.6em}{}
\renewcommand{\theparagraph}{\alph{paragraph}}
\makeatletter
\@addtoreset{paragraph}{section}
\@addtoreset{paragraph}{subsection}
\makeatother
\titleformat{\paragraph}[runin]{\normalfont\normalsize\itshape}{\theparagraph:}{0.4em}{}
\titlespacing*{\paragraph}{0pt}{1.0ex plus .2ex}{0.5em}
\titlespacing*{\section}{0pt}{1.6ex plus .3ex}{1.0ex}
\titlespacing*{\subsection}{0pt}{1.4ex plus .2ex}{0.8ex}
\titlespacing*{\subsubsection}{0pt}{1.2ex plus .2ex}{0.6ex}
\usepackage{newtxtext,newtxmath}   
\usepackage{amsmath}
\usepackage{graphicx,booktabs,microtype,bm,textcomp}
\usepackage{cite,url,xurl}
\usepackage[hidelinks]{hyperref}
\usepackage{authblk}
\usepackage{caption}
\newcommand{\capDelta}{\ensuremath{\Delta}}  

\title{PIDS-Bench: Evaluating Prompt-Injection Detectors Under Over-Defense,
Obfuscation, and Distribution Shift}
\author[1]{Yusuf Khalid Shire}
\author[1]{Sang-Chul Kim}
\affil[1]{Department of Software Convergence, Kookmin University, Seoul, Republic of Korea}
\date{}
\begin{document}
\twocolumn[
    \begin{@twocolumnfalse}
    \maketitle
    \begin{abstract}
    Prompt-injection detectors are typically evaluated using aggregate $F_1$ on in-distribution test data, which offers limited insight into behavior under distribution shift, particularly on the benign side of the decision boundary, where false positives impose direct operational cost yet are seldom measured. We present PIDS-Bench, a frozen multi-axis benchmark that jointly evaluates attack detection and benign false-positive behavior at fixed thresholds, spanning in-distribution inputs, hard-benign prompts that mimic injection structure without malicious intent, obfuscated attacks, and domain and structural distribution shifts. We evaluate seven detectors (learned baselines, external prompt-injection classifiers, and broad-safety comparators) alongside a rule-based lower-bound reference.
    Multi-axis evaluation exposes a failure mode that aggregate $F_1$ conceals. A detector exceeding $F_1 = 0.98$ on the held-out split still misclassifies roughly one-third of an externally-sourced benign subset drawn from public corpora and restricted to security-adjacent content. Across a full threshold sweep and five training seeds, no internal detector reaches an operating point satisfying $F_1 \geq 0.95$ and hard-benign $\mathrm{FPR} \leq 0.10$ together on this stress distribution. Decomposing by provenance, we find that hard-negative augmentation nearly eliminates over-defense on curated stress inputs but leaves it substantially intact on externally-sourced prompts, a pattern we term \emph{provenance-sensitive over-defense}. The asymmetry holds across both fine-tuned architectures and does not diminish as the augmentation pool grows, with the externally-sourced $\mathrm{FPR}$ remaining far above the $0.10$ target. Whether augmentation matched to the externally-sourced distribution would close this gap is untested; threshold calibration and curated-style augmentation alone do not.
    \end{abstract}
    \noindent\textbf{Keywords:} prompt injection, large language models, benchmark,
  adversarial robustness, over-defense, distribution shift.
  \vspace{1.5em}
  \end{@twocolumnfalse}
]

\renewcommand{\thefootnote}{}
\footnotetext{\raggedright Published in \textit{IEEE Access}, vol.~14, 2026.
\textcopyright~2026 The Authors. Licensed under CC BY 4.0.
DOI: \href{https://doi.org/10.1109/ACCESS.2026.3728186}{10.1109/ACCESS.2026.3728186}\\
Corresponding authors: khalidshire@kookmin.ac.kr, sckim7@kookmin.ac.kr}
\renewcommand{\thefootnote}{\arabic{footnote}}

\section{Introduction}
Large language models (LLMs) are increasingly deployed for document processing, code generation, customer interaction, and autonomous tool use. As these systems take on agent-like operation (multi-step reasoning, external tool invocation, interaction with sensitive data), prompt injection has become a practical security risk: adversarial text, supplied directly or retrieved from external content, attempts to override system instructions, redirect outputs, or induce disclosure of protected information~\cite{perez2022ignore, greshake2023not}. In agentic settings such inputs can propagate across reasoning steps and trigger unauthorized tool calls or data exposure.

A common defense places a binary prompt-injection detector in front of the model: inputs classified as malicious are blocked, the rest forwarded. This admits two error types with asymmetric, often unmeasured costs. False negatives let adversarial inputs through; false positives block legitimate queries (including security-related or instruction-like requests with no harmful intent) at a direct usability cost. In our evaluation, a detector reaching $F_1 > 0.98$ on held-out data still blocks roughly one-third of an externally-sourced, security-adjacent benign subset, rising further on a curated subset built to concentrate worst-case inputs. These rates are measured on stress distributions designed to surface over-defense, not to estimate deployment false-positive rates (FPR); even so, the cost they reveal is invisible under standard aggregate evaluation.

Most prompt-injection evaluations report a single aggregate metric, typically $F_1$, on a held-out split drawn from the training distribution. This measures separability under controlled conditions but misses deployment realities: benign inputs may share lexical or structural features with attacks, adversarial content may be obfuscated without changing intent, and input domain or format may shift between training and use. These concerns echo established robustness findings: held-out accuracy overstates reliability and structured stress tests are needed to surface concealed failures~\cite{ribeiro2020beyond}, evaluation across distributions can reorder model rankings~\cite{liang2022helm}, and strong in-distribution performance protects neither against distribution shift~\cite{hendrycks2020pretrained} nor surface-level perturbation~\cite{jin2020bert}. The principle is well established in general robustness evaluation but has not been applied systematically to binary gateway classifiers, where benign false positives are a primary operational cost.

Prior prompt-injection work has focused mainly on attack construction and end-to-end system compromise~\cite{schulhoff2023ignore}, with defenses evaluated largely in-distribution~\cite{inan2023llamaguard, wei2023jailbroken}. Existing multi-condition benchmarks (BIPIA~\cite{yi2023benchmarking}, InjecAgent~\cite{zhan2024injecagent}, AgentDojo~\cite{debenedetti2024agentdojo}, and HarmBench~\cite{mazeika2024harmbench}) evaluate the full pipeline under attack. End-to-end attack success, however, conflates detector behavior with downstream model compliance and does not decompose the two-sided error cost of a gateway classifier, whose benign false-positive rate carries a usability cost that attack-success metrics do not reflect. We isolate the detector-level setting and evaluate it systematically across seven detectors (three internal learned baselines, two external prompt-injection classifiers, and two broad-safety comparators) against a rule-based lower-bound reference.

Our protocol spans four complementary conditions: in-distribution inputs, hard-benign prompts that resemble attack structure without malicious intent, obfuscated attacks, and domain and structural distribution shift. These are realized through five frozen stress sets (Table~\ref{tab:splits}), under consistent threshold selection and split assignment within each model group. We operationalize it with a purpose-built dataset combining externally-sourced and synthesized inputs, with seed-family separation enforced to prevent train--evaluation overlap; the dataset instantiates the protocol and is not intended as a general-purpose corpus. Because model groups operate under different threshold configurations, all results are reported and interpreted within groups (Section~\ref{sec:protocol}).

\subsection*{Contributions}
\begin{itemize}
\item \textbf{Provenance-sensitive over-defense, a mitigation limit not previously reported.} Decomposing hard-benign false positives by the provenance of the benign input reveals an asymmetry that aggregate over-defense measures conceal: hard-negative augmentation nearly eliminates over-defense on curated inputs but leaves externally-sourced over-defense substantially intact, across both fine-tuned architectures and across augmentation pool sizes, at no measurable in-distribution cost (Table~\ref{tab:ablation}). We term this \emph{provenance-sensitive over-defense}. To our knowledge, no prior benchmark decomposes over-defense by input provenance or shows that a standard mitigation clears it on curated inputs yet fails to transfer to externally-sourced ones. A further decomposition, by injection-like surface structure and by source corpus, shows the effect is not merely structural: structure is the dominant driver and operates within every provenance, while source origin adds a smaller, statistically independent effect (Section~\ref{sec:confound}). Whether augmentation matched to the externally-sourced distribution would close the gap is untested; threshold calibration and curated-style augmentation alone do not.
\item \textbf{A threshold-frontier characterization of the detection/over-defense trade-off.} Building on evidence that prompt-injection detectors over-block benign inputs carrying security-adjacent language~\cite{injecguard, capture} and on calibration-focused low-FPR evaluation~\cite{promptshield}, we characterize the trade-off across the full operating range rather than at a single threshold. Sweeping $\tau \in [0,1]$ over five seeds per architecture, we find no operating point on the benchmark that meets $F_1 \geq 0.95$ and externally-sourced hard-benign $\mathrm{FPR} \leq 0.10$ together, across every architecture and seed evaluated. Even at each architecture's most favorable point, the minimum attainable externally-sourced FPR is $0.226 \pm 0.058$ for DeBERTa-v3-FT, $0.277 \pm 0.099$ for DistilBERT, and $0.241$ for term frequency--inverse document frequency (TF-IDF) with logistic regression. The validation-tuned threshold is itself highly unstable across seeds for the fine-tuned detectors (for DeBERTa-v3-FT, $\tau$ ranges from $0.01$ to $0.904$), yet the constraint is unmet at every threshold and seed. This pattern is consistent with a learned-feature account rather than a calibration artifact (Section~\ref{sec:sweep}); we state it as an interpretation, not an established mechanism.
\item \textbf{PIDS-Bench: a frozen, checksum-verified multi-axis benchmark.} PIDS-Bench measures attack detection and benign false-positive behavior jointly across in-distribution inputs, hard-benign prompts, obfuscated attacks (leetspeak, homoglyph, and zero-width transforms), and domain and structural distribution shift. It separates two failure modes that a single distribution-shift score conflates: attack-side recall collapse, where DistilBERT's code-domain recall falls to $0.435$ while remaining near-perfect elsewhere; and benign-side false-positive inflation, where TF-IDF reaches a legal-domain FPR of $0.252$ with recall unchanged. All splits are frozen, with SHA-256 checksums recorded in \texttt{FREEZE\_MANIFEST.txt} and released publicly for reproducible evaluation under a shared protocol.
\end{itemize}

\section{Related Work}

\subsection{Prompt-Injection Attacks and Threat Taxonomies}

Prompt injection as a named attack class was first studied systematically by Perez and Ribeiro~\cite{perez2022ignore}, who showed that adversarial text appended to user input can override model instructions and redirect LLM behavior in predictable ways. Greshake et al.~\cite{greshake2023not} extended this to the indirect setting, in which malicious instructions are concealed in externally retrieved content (web pages, documents, or API outputs) rather than supplied directly, establishing that an LLM-integrated system's attack surface includes any untrusted content the model processes. Schulhoff et al.~\cite{schulhoff2023ignore} documented the breadth of effective injection strategies through a large-scale prompt-hacking competition, providing an empirical characterization of attack diversity in deployed systems. Toyer et al.~\cite{toyer2023tensortrust} contributed Tensor Trust, a corpus of over 126{,}000 human-generated injection attacks and 46{,}000 defenses collected through an online game; unlike automatically generated adversarial prompts, these exhibit structured and compositional patterns, surfacing common failure modes such as instruction overriding and prompt leakage.

The threat is most consequential in agentic deployments, where a successful injection can propagate beyond the initial response and enable unauthorized tool calls or data exfiltration~\cite{greshake2023not, zhan2024injecagent}. PIDS-Bench is designed to evaluate gateway detectors in settings where both missed attacks and over-flagged benign inputs carry direct operational cost.

\subsection{Detection and Defense}

Defenses against adversarial prompt behavior fall into two broad categories. Alignment-based mitigation trains the underlying model to resist adversarial instructions, but such defenses can be systematically bypassed by constructed adversarial inputs~\cite{wei2023jailbroken}, motivating dedicated input-level classifiers as a more predictable enforcement layer. Within the classifier-based paradigm, Inan et al.~\cite{inan2023llamaguard} introduced Llama Guard and helped establish the gateway classifier as a common architectural pattern; subsequent releases extended this line with broader category taxonomies and multilingual coverage~\cite{metallamaguard2_2024, llamaguard3_1b_modelcard}, and WildGuard~\cite{han2024wildguard} introduced a three-way classifier distinguishing prompt harmfulness, response harmfulness, and refusal behavior. These systems target broad multi-category safety and are not specifically optimized for prompt-injection benign false positives. We include Llama Guard~3-1B as a broad-safety comparator to examine whether general-safety training transfers to prompt-injection detection. On the input-filtering side, Jain et al.~\cite{jain2023baseline} evaluated strategies such as perplexity-based filtering and paraphrase preprocessing, finding that no single approach offers robust protection across attack types.

Several purpose-built prompt-injection classifiers have been released as off-the-shelf components, including ProtectAI DeBERTa-v3-base-prompt-injection-v2~\cite{protectai2023deberta} and DeBERTa-PI~\cite{deepset2023deberta}. Their model cards report numbers on proprietary held-out sets that are neither frozen nor independently verifiable, and to our knowledge no shared, reproducible evaluation protocol covering these models has been published. PromptGuard~2 (86M)~\cite{meta2024promptguard} reports aggregate safety classification rather than prompt-injection metrics under a controlled threshold protocol. PIDS-Bench evaluates all three under a frozen, fixed-threshold protocol on a single shared evaluation set.

Recent work has begun to study over-defense directly at the prompt-injection detection layer. Li et al.~\cite{injecguard} introduced the NotInject dataset to quantify how guardrail models misclassify benign inputs containing injection-associated trigger words, attributing the behavior to lexical shortcut learning and proposing a token-level training-side mitigation; because NotInject's benign inputs are seeded with trigger words, its authors note that it may underestimate over-defense on realistic, domain-varied benign prompts. Kholkar and Ahuja~\cite{capture} proposed a context-aware benchmark that jointly probes attack detection and over-defense using small numbers of in-domain examples, reporting both elevated false negatives on adversarial inputs and elevated false positives on benign ones. Jacob et al.~\cite{promptshield} focused on the deployable low-false-positive-rate regime, arguing that the imbalance between benign and injection traffic makes a very low FPR a primary requirement for a fielded detector. Together, these works establish that benign over-blocking is a real and measurable property of prompt-injection detectors, yet to our knowledge none reports a mitigation that closes over-defense on externally-sourced, security-adjacent benign prompts drawn from public corpora, the regime PIDS-Bench measures directly.

PIDS-Bench builds on this line in three respects. First, rather than characterizing over-defense at fixed operating points, we trace it across the full threshold range and show that for every internal architecture evaluated here, no operating point jointly satisfies $F_1 \geq 0.95$ and hard-benign $\mathrm{FPR} \leq 0.10$ (Section~\ref{sec:sweep}); the trade-off is consistent with a learned-feature account rather than a calibration artifact alone. Second, we measure over-defense alongside attack recall, obfuscation robustness, and domain and structural distribution shift within a single frozen, checksum-verified protocol applied uniformly across internal, external, and broad-safety detectors. Third, whereas the over-defense sets in NotInject and CAPTURE are model-generated and seeded with trigger vocabulary, we additionally measure over-defense on externally-sourced, topic-filtered prompts drawn from public corpora, and find the effect present under both construction regimes. We refer to this asymmetry between curated and externally-sourced over-defense as \emph{provenance-sensitive over-defense}.

The mitigation we study, augmenting training with hard-negative benign examples that carry injection-associated features, draws on a broader body of work on data-centric robustness. Counterfactual data augmentation has been used to reduce reliance on spurious lexical correlations in text classification. Wang and Culotta~\cite{wang2021counterfactual} identify likely causal features and add counterfactual training examples that flip the label, producing classifiers that hold up under distribution shift where a standard classifier degrades sharply. That work shares both our diagnosis, that fine-tuned classifiers exploit surface features co-occurring with the label rather than the underlying concept~\cite{mccoy2019right}, and our remedy of targeted augmentation. Our finding is orthogonal to the choice of augmentation method: the benefit of curated hard-negative augmentation is provenance-dependent, clearing over-defense on inputs that resemble the augmentation distribution while leaving externally-sourced over-defense largely intact. Prior augmentation work reports aggregate robustness gains; to our knowledge, the asymmetry across input provenance, and its persistence across pool sizes, has not been reported.

Exaggerated safety behavior in aligned LLMs (where benign prompts are refused because they superficially resemble unsafe content) has been studied at the generation layer~\cite{rottger2024xstest}. The detector-layer setting is structurally distinct in two ways relevant to evaluation. The cost structure differs: a generation-layer refusal can sometimes be recovered through rephrasing or downstream reasoning, while a gateway block is terminal and the input never reaches the model. The decision surface also differs: exaggerated safety in generation is a function of training and decoding, not of a thresholded score, whereas detector-layer over-blocking exposes a continuous decision boundary that can be analyzed across operating points, a property we exploit in Section~\ref{sec:sweep} to show that the failure mode persists across the swept threshold range $\tau \in [0,1]$.

\subsection{Benchmarks for Prompt Injection and LLM Safety}

Existing benchmarks for LLM robustness share a property that limits their usefulness for gateway-detector evaluation: they measure the behavior of the full LLM pipeline under attack, not the behavior of an isolated binary classifier. BIPIA~\cite{yi2023benchmarking} benchmarks end-to-end indirect injection success across tool-use and retrieval scenarios; InjecAgent~\cite{zhan2024injecagent} extends this to tool-calling agents; AgentDojo~\cite{debenedetti2024agentdojo} provides a dynamic task environment for structured agentic pipelines; and HarmBench~\cite{mazeika2024harmbench} offers a multi-axis framework for harmful-behavior elicitation and adversarial robustness across LLM families. JailbreakBench~\cite{chao2024jailbreakbench} standardizes jailbreak evaluation through curated behaviors, unified scoring, and a reproducible pipeline. Liu et al.~\cite{liu2024formalizing} formalized prompt-injection attack types and benchmarked attack and defense effectiveness across diverse settings.

These end-to-end benchmarks do not treat benign false-positive rate as a primary metric, enforce a shared threshold protocol across heterogeneous classifier families, or provide frozen checksum-verified splits for detector-level reproducibility. PIDS-Bench complements them along an orthogonal axis: rather than characterizing attack or defense effectiveness through model output, it isolates the detector and measures its benign-side behavior under hard-benign stress and multiple forms of distribution shift, with both sides of the error distribution treated as primary.

\subsection{Evaluation Methodology and Robustness Under Distribution Shift}

The evaluation design of PIDS-Bench is grounded in prior work on the limitations of single-metric assessment. Ribeiro et al.~\cite{ribeiro2020beyond} showed that held-out accuracy overstates model robustness and that structured behavioral probes are needed to surface failure modes that aggregate metrics conceal. The HELM benchmark~\cite{liang2022helm} demonstrated that single-metric model rankings frequently reverse when evaluation is broadened across tasks and distributions. Research on robustness under distribution shift has further shown that strong in-distribution performance does not protect fine-tuned transformers against degradation on shifted inputs~\cite{hendrycks2020pretrained}, and that fine-tuned models are susceptible to surface-form perturbations that preserve semantic content~\cite{jin2020bert}, findings that directly motivate the structural-shift and obfuscation conditions in Section~\ref{sec:benchmark}. Together, these results motivate evaluation that separates in-distribution performance from targeted stress conditions, the organizing principle of PIDS-Bench.

\section{Threat Model and Scope}

\subsection{System Model}
We consider an LLM-integrated application in which untrusted input is routed through a binary prompt-injection detector before reaching a downstream language model. The detector is a stateless input-level classifier: it receives a single input string and produces either a binary decision or a continuous score thresholded to a binary output. Inputs classified as injections are blocked at the gateway; benign inputs are forwarded unmodified. The detector sees no downstream model output, keeps no state across requests, and has no feedback loop to the language model. This architecture is held fixed across all evaluated detectors and defines the evaluation boundary of this work.

\subsection{Threat Objective}
The attacker constructs an input that the detector classifies as benign while embedding instructions intended to influence the downstream model. We define attack success strictly as \emph{detector evasion}: the input receives a benign classification and is forwarded. This deliberately excludes end-to-end harm: whether an evaded input induces harmful behavior depends on the downstream model, its alignment, and the application context, all outside our scope. Isolating the detector in this way is what lets us measure both error directions, missed attacks and over-blocked benign inputs, as properties of the filter itself.

\subsection{Attacker Capabilities}
We adopt a static threat setting. The attacker draws inputs from a fixed distribution over the six injection strategy types defined by the attack taxonomy (Section~\ref{sec:benchmark}), receives no feedback from the detector, and does not adapt inputs to detector output. Adaptive and feedback-driven attackers are a distinct setting beyond the present evaluation.

\subsection{Defender Assumptions}
The evaluated detectors differ in architecture, training regime, and output type; implementation details appear in Section~\ref{sec:protocol}, and the distinctions relevant to threat modeling are summarized here. Internal learned detectors---TF-IDF with logistic regression, DistilBERT, and DeBERTa-v3-FT---are trained on the frozen PIDS-Bench training split, with thresholds fixed by validation-$F_1$ maximization before any stress-set evaluation. External prompt-injection detectors---the ProtectAI model and DeBERTa-PI---are run inference-only at their native decision boundaries, reflecting off-the-shelf deployment. PromptGuard~2 (86M) and Llama Guard~3-1B are included as broad-safety comparators whose objectives extend beyond prompt injection, also evaluated at their native boundaries. In every case the decision boundary is fixed before evaluation and not adjusted per evaluation set, and no detector uses downstream output as feedback or adapts online.

\subsection{Scope Boundaries}
The evaluation targets the input-filtering stage only. Detector evasion means an input passes the classifier; it does not imply the downstream model executes the attacker's intent, and end-to-end compromise rates are not measured. The benchmark is further restricted to single-turn, English-language inputs; multi-turn and multilingual attacks are distinct distribution shifts present in real traffic and are identified as future work (Section~\ref{sec:future}).

\section{Experimental Protocol}
\label{sec:protocol}

\subsection{Evaluated Detectors}
\label{sec:detectors}
We evaluate seven detectors across three groups: three internal learned baselines, two external prompt-injection classifiers, and two broad-safety comparators. A rule-based heuristic is reported separately as a lower-bound reference. The three internal learned models---TF-IDF with logistic regression (LR), DistilBERT, and DeBERTa-v3-FT---form the primary within-group comparison set;
external prompt-injection detectors and broad-safety classifiers are evaluated under their respective native operating conditions and reported within their own groups (Table~\ref{tab:external_results}).

\paragraph{Internal baselines.}
The internal group comprises three learned detectors---TF-IDF with logistic regression, DistilBERT, and DeBERTa-v3-FT---trained on the frozen PIDS-Bench training split, together with a rule-based detector reported separately as a deterministic lower-bound reference. The \emph{rule-based detector} consists of manually defined keyword and regular-expression patterns organized around four indicator categories derived from the attack taxonomy in Table~\ref{tab:subtypes}: instruction-override phrases, system-prompt extraction attempts, role-play and persona-adoption triggers, and encoding-based obfuscation markers. A prompt is classified as an injection if any pattern matches; no scoring or weighting is applied. The rule set is defined prior to evaluation and is not adjusted based on benchmark results, serving as a deterministic heuristic lower bound rather than a competitive baseline. The \emph{TF-IDF with logistic regression} model uses unigram and bigram features to capture surface-level lexical patterns. The \emph{DistilBERT}~\cite{sanh2019distilbert} and \emph{DeBERTa-v3-FT}~\cite{he2021debertav3} classifiers are neural sequence models of different capacity, both initialized from publicly available pretrained checkpoints and fine-tuned on the training split.

\paragraph{External prompt-injection detectors.}
The \emph{ProtectAI model}~\cite{protectai2023deberta} and \emph{DeBERTa-PI}~\cite{deepset2023deberta} are evaluated in inference-only mode using their publicly released checkpoints, without adaptation. Both are applied at their default decision boundary of $0.5$, to observe how detectors trained on independent data behave under the evaluation conditions defined by PIDS-Bench.

\paragraph{Broad-safety comparators.}
\emph{PromptGuard~2 (86M)}~\cite{meta2024promptguard} is a malicious-prompt classifier trained on a broader safety objective beyond prompt injection. \emph{Llama Guard~3-1B}~\cite{llamaguard3_1b_modelcard} is an LLM-based safety classifier trained on a general safety taxonomy. Both are included to examine whether detectors trained on general safety objectives transfer to the prompt-injection setting, and are evaluated at their default decision boundaries.

\subsection{Implementation and Training Details}
\label{sec:training}

All internal learned detectors are trained on a single NVIDIA A100 GPU under a pinned software environment (Python~3.10; \texttt{transformers}~4.40.0; \texttt{torch}~2.1.2+cu121; \texttt{tokenizers}~0.19.1; \texttt{datasets}~2.19.0; \texttt{scikit-learn}~1.3.2; \texttt{numpy}~1.26.4). The complete \texttt{pip freeze} output is released with the benchmark artifact.

The TF-IDF model uses unigram and bigram features (\texttt{ngram\_range}=(1,2)), retaining terms appearing in at least two documents and at most 95\% of documents. The classifier uses balanced class weights, L2 regularization ($C = 1.0$), and up to 500 optimization iterations.

DistilBERT is initialized from \texttt{distilbert-base-uncased} and fine-tuned for two epochs with batch size~16, learning rate $5 \times 10^{-5}$, and maximum sequence length 256. DeBERTa-v3-FT is initialized from \texttt{microsoft/deberta-v3-base} and fine-tuned for three epochs with effective batch size~16 (per-device batch~8 with gradient accumulation~2), learning rate $2 \times 10^{-5}$, maximum sequence length 512, and linear warmup over the first 6\% of training steps. For both models, the checkpoint with the highest validation $F_1$ is selected. Optimization uses AdamW~\cite{loshchilov2019adamw} with $\beta_1 = 0.9$, $\beta_2 = 0.999$, $\epsilon = 10^{-8}$, and weight decay~0.01. Hyperparameters follow commonly used configurations for sequence classification and are not extensively tuned.

\paragraph{Multi-seed protocol.}
Each internal architecture is trained five times under identical hyperparameters with distinct random seeds $\{13,42,123,2024,7777\}$ on the same cluster environment. All reported quantitative results are five-seed mean $\pm$ standard deviation (Tables~\ref{tab:internal_results}, \ref{tab:ablation}), with the threshold sweep (Section~\ref{sec:sweep}) computed per seed and aggregated. TF-IDF with logistic regression converges to identical fitted solutions across all five seeds under the fixed preprocessing and optimization configuration; its reported standard deviations are therefore zero by construction. In-distribution (IID) $F_1$ reproduces across seeds within $\pm 0.002$; IID benign FPR varies somewhat more (DistilBERT cross-seed std $0.0047$), and hard-benign metrics exhibit substantially larger cross-seed variability, as quantified by the reported standard deviations.

\paragraph{Hard-negative ablation.}
The hard-negative ablation (Section~\ref{sec:ablation}) follows the same five-seed protocol: for each fine-tuned detector, all ten runs (five baseline, five augmented) are trained on the cluster under the pinned environment above, differing only in the inclusion of the hard-negative training set, with all other hyperparameters unchanged. All experiments are implemented in PyTorch~\cite{paszke2019pytorch} using the Hugging Face Transformers library~\cite{wolf2020transformers}. The training and evaluation code is available in the released repository (Data Availability).

\subsection{Evaluation Protocol and Thresholding}
\label{sec:thresholding}

All probabilistic internal detectors follow the same thresholding procedure. Per seed, a decision threshold $\tau$ is selected by grid search over $[0,1]$ with step size $0.002$, maximizing $F_1$ on the validation split. The selected threshold is then fixed and applied to the held-out test split and all stress sets. Across the five seeds the resulting thresholds span nearly the entire range for DeBERTa-v3-FT ($\tau \in \{0.01, 0.108, 0.152, 0.444, 0.904\}$; mean $0.324$) and are similarly dispersed for DistilBERT ($\tau \in \{0.112, 0.252, 0.382, 0.610, 0.664\}$; mean $0.404$), whereas TF-IDF with logistic regression selects $\tau = 0.518$ in every run; the spread for the fine-tuned neural detectors is itself a finding of this work, examined further in Section~\ref{sec:sweep}. External detectors and PromptGuard~2 are evaluated at their default decision boundary of $0.5$.

The evaluation does not impose a shared calibration protocol across all model groups: internal models are tuned on the validation split, while external systems retain their native decision boundaries. This creates two distinct operating conditions and limits the interpretability of cross-group threshold-dependent comparisons.

External and broad-safety detectors are evaluated for three purposes: (i)~to demonstrate that the protocol applies to off-the-shelf components without adaptation; (ii)~to characterize the operating profile a practitioner would obtain by deploying these models at their native decision boundaries; and (iii)~through the diagnostic threshold sweep in Section~\ref{sec:results}, to test whether the over-defense and structural-shift failure modes observed in internal baselines are properties of the PIDS-Bench evaluation distribution itself or artifacts of internal-model training. All relative-performance claims are based on within-group comparisons.

Threshold-dependent metrics ($F_1$, recall, FPR) are interpreted within groups only, because cross-group differences may reflect threshold placement rather than model capability. Threshold-independent area under the receiver operating characteristic curve (ROC-AUC) provides a more stable basis for cross-group comparison of ranking behavior, subject to the caveat that AUC weights operating regions uniformly and does not reflect performance at any specific decision rule. Where cross-group AUC comparisons appear in the discussion, they are presented as comparisons of ranking behavior under the respective operating conditions, not as evidence of overall superiority. Within-group results are reported in Tables~\ref{tab:internal_results} and~\ref{tab:external_results}.

The threshold sweep in Section~\ref{sec:sweep} evaluates internal models across the full range $\tau \in [0,1]$ per seed, reporting test $F_1$, aggregate hard-benign FPR, externally-sourced hard-benign FPR, and curated hard-benign FPR at each step.

\subsection{Metrics and Statistical Analysis}

Metrics are defined separately for each evaluation setting.

\paragraph{IID test split.}
$F_1$ is the primary metric, with real-source $F_1$ emphasized to reduce the influence of template-derived examples and mixed (aggregate) $F_1$ reported as a secondary measure. Precision, recall, and ROC-AUC are reported for detectors with calibrated probability outputs, including the external prompt-injection detectors. Broad-safety classifiers (PromptGuard~2, Llama Guard~3-1B) are excluded from ROC-AUC because they do not expose a continuous score amenable to threshold sweeping at evaluation time.

\paragraph{Hard-benign stress set.}
False-positive rate is the primary metric, reported in aggregate ($n = 1{,}472$) and decomposed by provenance into externally-sourced ($n = 872$) and curated ($n = 600$). This set contains only benign examples; attack recall is undefined. Label quality for this set follows the two-annotator blind audit described in Section~\ref{sec:hardbenign}.

\paragraph{Obfuscated attacks stress set.}
Attack recall is the primary metric. This set contains only injection examples; benign FPR is undefined.

\paragraph{Domain-shift and structural-shift stress sets.}
Both $F_1$ and benign FPR are reported. For domain shift, benign FPR is the primary measure of distribution-shift effect, with recall reported to track content transfer. For structural shift, both metrics are reported to distinguish failure behaviors under structured input changes.

\paragraph{Statistical analysis.}
All reported point estimates are five-seed means, with standard deviations characterizing cross-seed variance from random initialization. Because seed variance does not capture the sampling uncertainty of a finite stress set, we additionally report example-level bootstrap confidence intervals~\cite{efron1994introduction} ($10{,}000$ resamples) for the headline metrics (Section~\ref{sec:stats}, Appendices~\ref{app:cis} and~\ref{app:ext_cis}) and a paired analysis of the augmentation effect across seeds (Table~\ref{tab:paired}). Threshold-dependent metrics are interpreted within model groups only, since the evaluation does not impose a shared calibration condition across all groups; we therefore do not conduct cross-group significance testing.

\section{Benchmark Construction and Design}
\label{sec:benchmark}

\subsection{Design Rationale}
PIDS-Bench is built on a single premise: evaluating a prompt-injection detector requires probing the failure modes that matter in deployment, not only those that are easy to quantify. Four design decisions follow.

First, split assignment is enforced at the seed-family level. Every paraphrase and obfuscated descendant of a source example is assigned to the same partition before the train/validation/test split is applied, eliminating the evaluation inflation that arises when structurally similar examples straddle training and evaluation data~\cite{ribeiro2020beyond}.

Second, the held-out test split is kept separate from five purpose-built stress sets, each targeting a distinct failure mode. Evaluation reports four conditions: in-distribution, hard-benign, obfuscated, and distribution-shift. The latter three are realized through four stress sets (Table~\ref{tab:splits}): hard-benign, obfuscated-attacks, and the domain-shift and structural-shift sets. A fifth stress set, balanced-subtype, provides a per-subtype diagnostic disaggregation of the in-distribution condition rather than a separate stress axis. In-distribution $F_1$ and stress-set results are reported on separate axes because, as Section~\ref{sec:results} shows, collapsing them into a single score conceals qualitatively different failure patterns.

Third, benign false-positive rate is treated as a primary metric, on equal footing with attack detection. A detector that systematically over-blocks legitimate queries imposes direct operational cost: reduced system utility and disruption of the very security-adjacent workflows it is meant to protect. The hard-benign stress set is constructed to make this failure mode measurable. Treating benign false positives as a first-class criterion alongside missed detections follows prior practice in real-world safety moderation~\cite{markov2023holistic}.

Fourth, all splits are frozen, with SHA-256 checksums recorded in \texttt{FREEZE\_MANIFEST.txt} and distributed with the benchmark artifact.\footnote{The benchmark (split files, stress sets, and SHA-256 checksums) is available at \url{https://github.com/ShirePyDev/Prompt-Injection-Detection-System/tree/v1.0-pids-bench/data/pids_bench_v3}.} Freezing ties every reported result to a fixed, independently verifiable artifact and makes any post-hoc split adjustment detectable.

\subsection{Data Sources and Attack Taxonomy}
\label{sec:taxonomy}

\subsubsection{Injection Sources}
Injection examples are assembled from three public prompt-injection datasets and a set of template-derived attacks. The SPML Chatbot Prompt Injection dataset~\cite{spml2024dataset} is the largest external source, contributing 8{,}423 examples across the six attack subtypes, complemented by smaller sets from Qualifire~\cite{qualifire2023} (676) and deepset~\cite{deepset2023dataset} (126), both drawn from real deployment contexts. Template-derived attacks supply the remainder, concentrated in the two subtypes for which external labeled examples are scarce. The counts reported here are the final per-source totals in the released corpus, after deduplication and class-balanced sampling, and are fixed in the freeze manifest. Following standard practice for combining independently curated corpora~\cite{yi2023benchmarking, mazeika2024harmbench}, source labels are accepted as originally annotated without re-labeling.

\subsubsection{Benign Sources}
Benign examples are drawn from four public corpora: Stanford Alpaca~\cite{taori2023alpaca}, ChatBot Instructions~\cite{espejel2023chatbot}, OpenAssistant (OASST1)~\cite{oasst1}, and Databricks Dolly-15K~\cite{databricks_dolly_15k}, supplemented by template-generated benign prompts constructed alongside the injection templates. No corpus text is carried over verbatim: each seed is rewritten by GPT-4o-mini~\cite{openai2024gpt4omini} over one to three paraphrase rounds under a meaning-preserving prompt, and every released benign row is a paraphrase variant that inherits its seed's label. Across the combined train, validation, and test splits, the benign pool contains 16{,}637 rows: Alpaca (7{,}952), ChatBot Instructions (4{,}961), OpenAssistant (1{,}317), Dolly-15K (576), and 1{,}831 template-generated prompts. The exact per-split, per-source counts are fixed in the freeze manifest.
Benign labels are accepted as originally curated and, unlike the hard-benign stress set (Section~\ref{sec:hardbenign}), were not subject to independent inter-annotator audit at the same depth. Exact-match and near-duplicate filtering reduces leakage, but residual label noise or incidental injection-adjacent phrasing may remain in the in-distribution benign distribution. This qualifies the interpretation of in-distribution benign FPR (Table~\ref{tab:internal_results}), which reflects detector behavior on paraphrased, labels-as-curated text rather than on audited original-source text; Section~\ref{sec:limitations} discusses this further.

\subsubsection{Template Construction}
For the two attack subtypes with the fewest externally-sourced labeled examples, encoded attacks (\texttt{encoded\_attack}) and tool injection (\texttt{tool\_injection}), we supplement the corpus with template-derived attacks: GPT-4o-mini~\cite{openai2024gpt4omini} generates three surface-level paraphrases per real seed at sampling temperature 0.9, under a prompt that varies phrasing while preserving adversarial intent. Each variant inherits its parent's label and subtype annotation, so no new labeling decision is introduced. Template-derived rows account for 934 of the 3{,}918 held-out test examples (23.84\%) and 801 of the 2{,}070 test injections (38.70\%). Because these two subtypes are lexically more uniform than externally-sourced attacks, they may under-represent real-world surface diversity, a caveat that applies to every model evaluated on the test split.

\begin{table}[h]
\caption{Injection attack subtype definitions and proportions in the main injection pool (train+validation+test combined; $n = 18{,}363$ injection examples).}
\label{tab:subtypes}
\centering
\footnotesize
\resizebox{\columnwidth}{!}{%
\begin{tabular}{p{2.8cm}p{4.8cm}c}
\toprule
Subtype & Definition & \% \\
\midrule
\texttt{direct\_override} & Explicit instruction to disregard prior directives and adopt new behavior & 29 \\[4pt]
\texttt{contextual\_manip.} & Constructs false context to redirect model behavior without an explicit override & 24 \\[4pt]
\texttt{roleplay\_attack} & Uses persona adoption or fictional framing to bypass instruction constraints & 15 \\[4pt]
\texttt{data\_exfiltration} & Attempts to elicit protected content or system context from the model & 13 \\[4pt]
\texttt{tool\_injection} & Embeds injection content within tool-call parameters or API response fields & 10 \\[4pt]
\texttt{encoded\_attack} & Encodes malicious instructions via Base64, ROT13, or Unicode substitution & 9 \\
\bottomrule
\end{tabular}%
}
\end{table}

The injection class is partitioned into six subtypes, each capturing a distinct adversarial strategy. Assignment is deterministic: a priority-ordered rule matcher tests \texttt{direct\_override} first, then \texttt{roleplay\_attack}, \texttt{contextual\_manip.}, \texttt{data\_exfiltration}, \texttt{encoded\_attack}, and \texttt{tool\_injection}, placing each prompt in the highest-priority category it matches. The subtypes are therefore mutually exclusive by construction rather than by semantic boundary, and a single prompt may exhibit features of several. To gauge the reliability of these assignments, two graduate annotators independently labeled a stratified 150-row sample against the Table~\ref{tab:subtypes} definitions, yielding Cohen's $\kappa = 0.758$~\cite{cohen1960} (raw agreement 0.807; substantial on the Landis--Koch scale~\cite{landis1977}). Because subtype labels serve only diagnostic disaggregation and never enter detection training, residual disagreement does not affect the $F_1$ and FPR results in Section~\ref{sec:results}.

\subsection{Corpus Composition and Split Strategy}
\label{sec:composition}
The main corpus comprises 35{,}000 rows, partitioned into training (27{,}093), validation (3{,}989), and held-out test (3{,}918). The five stress sets add 8{,}172 further evaluation instances, for a total of 43{,}172. Of these, the obfuscated and balanced-subtype sets are built from held-out test prompts (Sections~\ref{sec:balanced} and~\ref{sec:distshift}), so 2{,}693 instances reuse text already in the test split (405 obfuscated and 2{,}288 balanced-subtype), and the benchmark therefore contains 40{,}479 unique labeled prompts. Table~\ref{tab:overlap} details this structure. The hard-benign, domain-shift, and structural-shift sets are independently sourced and disjoint from every other partition, whereas the obfuscated and balanced-subtype sets are deliberately derived from test prompts and reported as transformed evaluation instances rather than as new examples. Table~\ref{tab:splits} summarizes class balance and source composition across all partitions.

\begin{table}[h]
\centering
\caption{Benchmark composition and overlap with the held-out test split. ``Unique'' counts prompts whose text appears in no earlier partition; ``From test'' counts rows whose text is drawn from the test split (transformed or re-sampled). The hard-benign, domain-shift, and structural-shift stress sets are independently sourced and disjoint from all other partitions; the obfuscated and balanced-subtype sets are derived from held-out test prompts and are therefore evaluation instances rather than new unique prompts. Counts are computed by exact text matching across all partitions of the released artifact.}
\label{tab:overlap}
\setlength{\tabcolsep}{6pt}
\renewcommand{\arraystretch}{1.15}
\begin{tabular}{@{}lrrl@{}}
\toprule
Partition & Rows & From test & Status \\
\midrule
\multicolumn{4}{@{}l}{\textit{Main corpus}} \\
\quad Training         & 27{,}093 & 0   & unique \\
\quad Validation       & 3{,}989  & 0   & unique \\
\quad Test             & 3{,}918  & --- & reference \\
\addlinespace[2pt]
\multicolumn{4}{@{}l}{\textit{Stress sets}} \\
\quad Hard-benign      & 1{,}472  & 0       & unique \\
\quad Domain shift     & 2{,}000  & 0       & unique \\
\quad Structural shift & 1{,}998  & 0       & unique \\
\quad Obfuscated       & 405      & 405     & from test \\
\quad Balanced subtype & 2{,}297  & 2{,}288 & from test \\
\midrule
\textbf{Total instances} & \textbf{43{,}172} & 2{,}693 & \\
\textbf{Unique prompts}  & \textbf{40{,}479} &         & \\
\bottomrule
\end{tabular}
\end{table}

\begin{table}[h]
\centering
\caption{PIDS-Bench corpus and stress-set statistics. The Inj./Ben.\ column shows the injection-to-benign percentage split within each partition. Ext-src and Template columns show the proportion of rows drawn from externally-sourced datasets versus GPT-4o-mini paraphrase generation, used for data construction only; both columns apply to the main-corpus partitions alone and are undefined for the stress sets, where an em-dash marks the inapplicable field. The Total row counts evaluation instances; the obfuscated and balanced-subtype sets re-use held-out test prompts, so the benchmark contains 40{,}479 unique labeled prompts (Table~\ref{tab:overlap}).}
\label{tab:splits}
\small
\setlength{\tabcolsep}{4pt}
\resizebox{\columnwidth}{!}{%
\begin{tabular}{lrccc}
\toprule
Partition & Rows & Inj./Ben. & Ext-src \% & Template \% \\
\midrule
\multicolumn{5}{l}{\textit{Main corpus}} \\[2pt]
Training   & 27{,}093 & 53/47 & 66.4 & 33.6 \\
Validation & 3{,}989 & 51/49 & 76.8 & 23.2 \\
Test       & 3{,}918 & 53/47 & 76.2 & 23.8 \\
\midrule
\multicolumn{5}{l}{\textit{Stress sets (held-out only)}} \\[2pt]
Hard-benign        & 1{,}472 & 0/100 & --- & --- \\
Obfuscated attacks & 405 & 100/0 & --- & --- \\
Balanced subtype   & 2{,}297 & 48/52 & --- & --- \\
Domain shift       & 2{,}000 & 50/50 & --- & --- \\
Structural shift   & 1{,}998 & 50/50 & --- & --- \\
\midrule
\textbf{Total} & \textbf{43{,}172} & --- & --- & --- \\
\bottomrule
\end{tabular}%
}
\end{table}

Template-derived examples are concentrated in the training split so that evaluation preserves external provenance. The validation split serves one purpose (per-seed threshold selection for the internal learned baselines), while external classifiers are evaluated at their native decision boundaries without validation tuning. Section~\ref{sec:protocol} gives the complete threshold protocol.

\subsection{Balanced-Subtype Stress Set}
\label{sec:balanced}
The balanced-subtype stress set contains 2{,}297 rows, 2{,}288 of them drawn from the held-out test split (Table~\ref{tab:overlap}). It exposes per-subtype disparities that aggregate test $F_1$ conceals: a detector strong on frequent subtypes but weak on rarer ones looks uniform on the main split yet reveals the gap here. Subtype assignments follow the inter-annotator validation reported in Section~\ref{sec:taxonomy}, so the per-subtype results in Section~\ref{sec:results} are read as empirically grounded diagnostic groupings.
Constructed after paraphrase generation and split assignment, the set contains no training-family example. It targets 200 injections per subtype against 1{,}200 benign prompts drawn predominantly from the test-split benign pool; the injection side reaches 1{,}097 rows, short of the full quota because encoded attacks and tool injection fall below 200 after deduplication.

\subsection{Hard-Benign Stress Set}
\label{sec:hardbenign}
The hard-benign stress set contains 1{,}472 benign-only prompts designed to challenge over-defensive classifiers. Each discusses security concepts, prompt-injection mechanisms, system-prompt design, or model behavior in a descriptive or educational register, sharing substantial lexical overlap with injections while carrying no override instruction, request for protected content, or directive to alter system behavior. Representative categories include security documentation, vulnerability explanations, policy reasoning, and benign role-play.

\paragraph{Curation process.}
The 600 curated examples were authored under fixed guidelines that require injection-adjacent surface features (system-prompt references, instruction-like phrasing, security-related discussion) while excluding any manipulative intent. Each was reviewed before inclusion and deduplicated against the main corpus.

\paragraph{Label quality.}
A stratified sample of 200 hard-benign rows was audited under a three-category rubric: \textit{injection} (an explicit attempt to override, redirect, or manipulate the model), \textit{security-adjacent benign} (discussion of security or AI topics with no override instruction), or ordinary \textit{benign} content. The first author confirmed 197 of the 200 rows as benign and reclassified 3 as not clearly benign; 129 were additionally marked security-adjacent, indicating that the set predominantly exercises the intended stress condition. To assess reliability, a second annotator independently re-labeled all 200 rows under the same rubric, blind to the original labels and to every detector's predictions. Agreement on the benign--non-benign judgment that underlies the false-positive measurement was near-perfect, at 199 of 200 rows (Cohen's $\kappa = 0.85$, almost-perfect on the Landis--Koch scale~\cite{landis1977}); the lone disagreement was a content-moderation refusal string of genuinely ambiguous status. It was left under its original label; with only one row in dispute, no formal adjudication was required. Agreement on the finer security-adjacent designation was moderate ($\kappa = 0.53$, raw agreement $75.5\%$), the disagreements being largely one-directional, with the first annotator applying a broader threshold than the second. This pattern, an almost-perfect benign judgment alongside a merely moderate adjacency judgment, is consistent with these prompts occupying a genuinely ambiguous security-adjacent region, the property that makes them an effective over-defense stress set. Both annotators' labels are released in \texttt{hard\_benign\_audit\_sample.csv}. Because the benign judgment that determines the false-positive rate is reproduced under blind re-annotation, the over-defense finding does not rest on single-annotator label precision. The single benign--non-benign disagreement corresponds to an observed label-error rate of $0.5\%$ (1 of 200), with a $95\%$ Wilson interval of $[0.001, 0.028]$, an upper bound well below the externally-sourced false-positive rates this set is used to measure. Any residual mislabeling in this benign-only set could, moreover, only inflate the measured false-positive rate, never deflate it: a genuine injection mislabeled as benign is counted as a false positive precisely when a detector flags it, so contamination cannot render the reported rate a conservative underestimate.

\paragraph{Representativeness and scope.}
The externally-sourced subset (872 rows) is drawn from three public instruction and conversation corpora: LMSYS-Chat-1M~\cite{lmsys}, OpenAssistant (OASST1)~\cite{oasst1}, and Databricks Dolly-15K~\cite{databricks_dolly_15k}. Candidates were retained by a topical filter (content referencing security, system prompts, instructions, prompt design, or related model-behavior topics), so a prompt is included for its subject matter, not because any detector flagged it. Within each corpus we separate a \emph{plain} slice, matching the filter through general instruction-like or security-related phrasing, from an \emph{AI-adjacent} slice, whose prompts additionally discuss model behavior, prompting, or guardrail concepts; the split lets us distinguish over-defense driven by instruction structure from over-defense driven by AI-security subject matter. The curated subset (600 rows) was authored separately under the guidelines above, organized into eight equally sized families that each concentrate a distinct class of injection-adjacent benign language. Table~\ref{tab:hb_sources} gives the full per-source composition. Of the three externally-sourced corpora, only LMSYS is absent from benign training entirely; OpenAssistant and Dolly seeds appear in training as GPT-4o-mini paraphrases, but the hard-benign evaluation prompts are disjoint from the training rows, with verified zero text overlap. No hard-benign evaluation prompt was present during training. Because the evaluation rows were never seen during training (and, for LMSYS, the source corpus is entirely absent from training), over-blocking on this subset cannot reflect memorization of the specific benign examples the detectors were trained on. We return to this point in Section~\ref{sec:confound}.

 \begin{table}[h]
\centering
\caption{Per-source composition of the hard-benign stress set ($n=1{,}472$). The externally-sourced subset ($n=872$) is drawn from three public corpora, each split into a plain and an AI-adjacent slice; the curated subset ($n=600$) comprises eight equally sized authored families. Counts are fixed in the released artifact and recorded in the freeze manifest.}
\label{tab:hb_sources}
\setlength{\tabcolsep}{6pt}
\begin{tabular}{@{}llr@{}}
\toprule
Source & Slice / family & Rows \\
\midrule
\multicolumn{3}{@{}l}{\textit{Externally-sourced} ($n=872$)} \\
\addlinespace[1pt]
LMSYS-Chat-1M & plain       & 298 \\
              & AI-adjacent & 366 \\
OpenAssistant & plain       & 40  \\
              & AI-adjacent & 100 \\
Dolly-15K     & plain       & 18  \\
              & AI-adjacent & 50  \\
\midrule
\multicolumn{3}{@{}l}{\textit{Curated} ($n=600$)} \\
\addlinespace[1pt]
Authored families & Security education         & 75 \\
                  & Red-team framing           & 75 \\
                  & Attack analysis            & 75 \\
                  & Guardrail design           & 75 \\
                  & Policy / meta-discussion    & 75 \\
                  & Imperative benign          & 75 \\
                  & Fiction / role-play benign & 75 \\
                  & Technical documentation    & 75 \\
\bottomrule
\end{tabular}
\end{table}
 
Aggregate FPR figures computed over the full 1{,}472-row set therefore characterize detector behavior on a stress distribution of security-adjacent benign inputs, not the expected false-positive rate across arbitrary deployment traffic. The per-source decomposition in Table~\ref{tab:hardbenign_origin} distinguishes the two regimes: externally-sourced FPR reflects behavior on topic-filtered samples from public corpora, while curated FPR reflects behavior on constructed worst-case inputs. Neither figure alone predicts deployment FPR without knowledge of the base rate of security-adjacent language in the target application.

\subsection{Distribution-Shift Stress Sets}
\label{sec:distshift}
The benchmark separates two forms of distribution shift onto distinct axes, because they expose different failures: domain shift surfaces content-level generalization gaps, while structural shift is dominated by benign-side false-positive inflation from surface-form associations learned in training.

\paragraph{Domain shift.}
The domain-shift set contains 2{,}000 rows across four specialized domains (medical, legal, finance, and code), each with 250 benign and 250 injection examples. The benign side, drawn from four external corpora used nowhere else in the benchmark (MedQuAD, the European Court of Human Rights (ECtHR) subset of LexGLUE, Financial-QA-10K, and CodeSearchNet), tests whether a detector trained on general text holds a low false-positive rate as vocabulary shifts. The injection side is generated with GPT-4o-mini under a fixed prompt that phrases an attack in each domain's vocabulary. Both sides are near-duplicate filtered against training (character $n$-gram TF-IDF, cosine threshold 0.90), leaving the set disjoint from all main-corpus partitions (Table~\ref{tab:overlap}). Because the injections are newly generated rather than resampled, attack-side results measure generalization to domain-shifted phrasing rather than memorized content. Benign FPR is the primary measure here, attack recall secondary.

\paragraph{Per-domain analysis.}
Aggregate metrics hide domain-specific behavior (Table~\ref{tab:domain_ood_breakdown}). The two architectures fail in opposite ways. DistilBERT's attack recall collapses in the code domain (0.435) while its false-positive rate stays low, a failure on the attack side. In the legal domain the pattern inverts: recall stays near-perfect but the false-positive rate rises, blocking legitimate queries. That legal-domain rate carries wide cross-seed variance ($0.254 \pm 0.198$), so we read it as evidence of over-blocking in this domain rather than as a precise estimate. TF-IDF with logistic regression keeps higher code recall (0.916) but shows the same elevated legal FPR (0.252), deterministically. Robustness under domain shift is thus not one property: the failure mode depends on domain vocabulary and structure.

\begin{table}[h]
\centering
\caption{Per-domain breakdown of domain-shift performance for DistilBERT (five-seed mean $\pm$ std) and TF-IDF with logistic regression (deterministic); 250 benign and 250 injection examples per domain. All values at $\tau = 0.5$.}
\label{tab:domain_ood_breakdown}
\small
\resizebox{\columnwidth}{!}{%
\begin{tabular}{lcccc}
\toprule
& \multicolumn{2}{c}{DistilBERT} & \multicolumn{2}{c}{TF-IDF + LR} \\
\cmidrule(lr){2-3}\cmidrule(lr){4-5}
Domain & FPR & Recall & FPR & Recall \\
\midrule
Medical & $0.0016 \pm 0.0022$ & $0.9992 \pm 0.0018$ & 0.000 & 0.988 \\
Legal   & $0.254 \pm 0.198$ & $0.992 \pm 0.018$ & 0.252 & 0.968 \\
Finance & $0.010 \pm 0.009$ & $1.000 \pm 0.000$ & 0.012 & 0.996 \\
Code    & $0.029 \pm 0.022$ & $0.435 \pm 0.151$ & 0.004 & 0.916 \\
\bottomrule
\end{tabular}%
}
\end{table}

\paragraph{Structural shift.}
The structural-shift set contains 1{,}998 rows: three surface-form transforms applied to 333 benign and 333 injection seeds (666 in total), each transformed three ways. Benign seeds come from the hard-benign set (Section~\ref{sec:hardbenign}); injection seeds are GPT-4o-mini-generated and near-duplicate filtered against training, so no transformed variant appears in any main-corpus partition (Table~\ref{tab:overlap}). JSON wrapping (\texttt{json\_wrap}) embeds the prompt as a JSON value, mimicking tool-call output; prompt dilution (\texttt{prompt\_dilute}) pads the prompt with surrounding text, mimicking injection buried in a retrieved passage; instruction-prefix mimicry (\texttt{instruction\_prefix\_mimic}) prepends a system-prompt-like prefix, probing whether authoritative formatting misleads the detector. The transforms preserve semantic intent by design (checked during construction but not independently annotated), so instruction-prefix mimicry, the most likely to shift a borderline label, warrants the most caution. Because its benign seeds already carry security-adjacent language, it tests whether authoritative formatting compounds existing over-defense rather than introducing new injection-adjacent content. Results characterize sensitivity to these specific constructions, not to arbitrary format variation (Section~\ref{sec:limitations}).

\paragraph{Held-out obfuscation evaluation.}
To probe surface-level evasion, we collect the 405 held-out injection examples carrying a character-level obfuscation: \textit{leetspeak} keyword substitution, \textit{homoglyph} replacement, or \textit{zero-width} insertion. These transform families already appear in training, so the set measures sensitivity to known perturbations rather than robustness to novel evasion (Section~\ref{sec:limitations}).

\section{Results}
\label{sec:results}
\subsection{Main Multi-Axis Results}
Table~\ref{tab:internal_results} reports the three internal baselines as five-seed mean $\pm$ standard deviation at a fixed threshold $\tau = 0.5$, the operating point at which cross-seed averaging is directly comparable. Table~\ref{tab:external_results} reports the external and broad-safety detectors, which are not trained on PIDS-Bench data and run under their own decision rules; the tables share column axes for readability, but threshold-dependent metrics are not compared row-by-row across them, since the groups operate under different thresholding regimes.

\begin{table*}[t]
\caption{Multi-axis evaluation for internal baselines, reported as five-seed mean $\pm$ standard deviation at fixed threshold $\tau = 0.5$. hb-FPR is the aggregate over the full hard-benign set ($n = 1{,}472$); the externally-sourced ($n = 872$) and curated ($n = 600$) decomposition is given in Table~\ref{tab:hardbenign_origin}. TF-IDF + LR converges to identical solutions across seeds and is reported without variance. $\uparrow$ higher is better; $\downarrow$ lower is better.}
\label{tab:internal_results}
\centering
\setlength{\tabcolsep}{4pt}
\resizebox{\textwidth}{!}{%
\begin{tabular}{llcccccccc}
\toprule
Category & Model
& IID F1$\uparrow$
& Real-source F1$\uparrow$
& IID Ben-FPR$\downarrow$
& hb-FPR$\downarrow$
& Obf.\ Recall$\uparrow$
& Domain-shift F1$\uparrow$
& Struct.-shift FPR$\downarrow$
& ROC-AUC$\uparrow$ \\
\midrule
Classical ML
& TF-IDF + LR
& 0.9656
& 0.9467
& 0.0514
& 0.4090
& 0.9654
& 0.9508
& 0.7918
& 0.9941 \\
Internal FT
& DistilBERT
& $0.9831 \pm 0.0017$
& $0.9738 \pm 0.0026$
& $0.0245 \pm 0.0047$
& $0.4702 \pm 0.0823$
& $0.9877 \pm 0.0058$
& $0.8886 \pm 0.0214$
& $0.7826 \pm 0.0963$
& $0.9974 \pm 0.0002$ \\
Internal FT
& DeBERTa-v3-FT
& $0.9882 \pm 0.0008$
& $0.9825 \pm 0.0009$
& $0.0124 \pm 0.0013$
& $0.3825 \pm 0.0474$
& $0.9867 \pm 0.0079$
& $0.9604 \pm 0.0099$
& $0.8108 \pm 0.0617$
& $0.9983 \pm 0.0004$ \\
\bottomrule
\end{tabular}%
}
\end{table*}

\paragraph{Internal baselines.}
DeBERTa-v3-FT leads the internal baselines in-distribution (IID $F_1$ $0.9882 \pm 0.0008$, domain-shift $F_1$ $0.9604 \pm 0.0099$, ROC-AUC $0.9983 \pm 0.0004$), but this strength does not carry to benign-side behavior under stress. At $\tau = 0.5$ it blocks $0.3144 \pm 0.0525$ of externally-sourced security-adjacent benign queries, roughly one in three, with aggregate hard-benign FPR $0.3825 \pm 0.0474$, curated-subset FPR $0.4813 \pm 0.0855$ (Table~\ref{tab:hardbenign_origin}), and structural-shift FPR $0.8108 \pm 0.0617$. Later sections take the externally-sourced hard-benign FPR as the primary benign-side measure, since it reflects topic-filtered prompts from public corpora rather than constructed worst-case inputs. DistilBERT reaches IID $F_1$ $0.9831 \pm 0.0017$ under the same procedure with a higher aggregate hard-benign FPR ($0.4702 \pm 0.0823$), and TF-IDF with logistic regression, which converges to one solution across seeds, attains IID $F_1$ $0.9656$ with aggregate hard-benign FPR $0.4090$ and structural-shift FPR $0.7918$. The elevated hard-benign FPR is thus neither architecture-specific nor confined to neural models: all three internal detectors over-block security-adjacent benign inputs.

The rule-based detector, included only as a lower-bound reference, is telling despite its simplicity. On the IID split it is highly conservative, firing on almost nothing (precision $1.0$, recall $0.035$, $F_1 = 0.067$), yet its keyword and regular-expression patterns flag $0.229$ of the hard-benign set (337 of $1{,}472$). A deterministic matcher that virtually never fires on ordinary text still over-blocks security-adjacent benign inputs at this rate, which locates the source of over-blocking in security-topic vocabulary itself rather than solely in learned feature associations.

\paragraph{Validation-tuned operating points.}
Table~\ref{tab:internal_results} fixes $\tau = 0.5$ for comparability, but in deployment each model would operate at its validation-$F_1$-maximizing threshold. For the fine-tuned detectors these thresholds are strikingly unstable across seeds ($0.01$ to $0.904$ for DeBERTa-v3-FT, $0.112$ to $0.664$ for DistilBERT), whereas the linear baseline selects $\tau = 0.518$ in every run. Section~\ref{sec:sweep} takes up this instability and its consequences.

\paragraph{External and broad-safety detectors.}
The external and broad-safety detectors are evaluated at their default boundary ($\tau = 0.5$), the configuration a practitioner obtains without adaptation, with threshold-independent ROC-AUC reported where continuous scores are available ($0.9741$ for ProtectAI, $0.8712$ for DeBERTa-PI; omitted for PromptGuard~2 and Llama Guard~3-1B, which emit categorical verdicts). To separate intrinsic failure from threshold placement, we also sweep each probabilistic external detector on the validation split under the same validation-$F_1$ rule and report a second, diagnostic row. The selected thresholds are extreme ($\tau = 0.002$ for ProtectAI, $\tau = 0.986$ for DeBERTa-PI), and neither removes the elevated structural-shift FPR seen at the default boundary; DeBERTa-PI flags every structural-shift benign input ($\mathrm{FPR} = 1.000$) at both. Threshold-dependent comparisons are therefore read within groups.

The two broad-safety classifiers act as weak injection detectors without adaptation: PromptGuard~2 and Llama Guard~3-1B reach IID $F_1$ of only $0.539$ and $0.569$, recalling under half of injections. Their hard-benign FPRs ($0.152$, $0.180$) are nonetheless \emph{lower} than the injection-trained internal detectors' ($0.383$ for DeBERTa-v3-FT, $0.470$ for DistilBERT). This is consistent with over-defense being most pronounced in detectors trained specifically on injection data rather than a generic property of safety classifiers.

\begin{table*}[t]
\caption{External and broad-safety detectors. External prompt-injection detectors are evaluated at their default boundary ($\tau = 0.5$) and at a threshold selected by sweeping the PIDS-Bench validation split under the validation-$F_1$ rule; the swept rows are diagnostic and are not controlled comparisons with internally trained models. Broad-safety models are reported at their native operating points. ROC-AUC is shown for detectors exposing calibrated continuous scores. Llama Guard~3-1B was evaluated in inference-only mode under a separate inference environment. $\uparrow$ higher is better; $\downarrow$ lower is better.}
\label{tab:external_results}
\centering
\setlength{\tabcolsep}{4pt}
\resizebox{\textwidth}{!}{%
\begin{tabular}{llccccccc}
\toprule
Category & Model ($\tau$)
& IID F1$\uparrow$ & IID Recall$\uparrow$ & hb-FPR$\downarrow$
& Obf.\ Recall$\uparrow$ & Domain-shift F1$\uparrow$ & Struct.-shift FPR$\downarrow$ & ROC-AUC$\uparrow$ \\
\midrule
External PI & ProtectAI ($\tau{=}0.5$)    & 0.9036 & 0.858 & 0.216  & 0.906 & 0.9103 & 0.4525 & 0.9741 \\
External PI & ProtectAI ($\tau{=}0.002$)  & 0.9285 & 0.913 & 0.322  & 0.965 & 0.9617 & 0.6647 & 0.9741 \\
External PI & DeBERTa-PI ($\tau{=}0.5$)   & 0.8044 & 0.944 & 0.804  & 0.970 & 0.7950 & 1.000  & 0.8712 \\
External PI & DeBERTa-PI ($\tau{=}0.986$) & 0.8133 & 0.887 & 0.694  & 0.943 & 0.7973 & 1.000  & 0.8712 \\
\midrule
Broad-safety & PromptGuard~2 (86M) & 0.5390 & 0.369 & 0.1515 & 0.3235 & 0.1900 & 0.3323 & --- \\
Broad-safety & Llama Guard~3-1B    & 0.5692 & 0.420 & 0.1800 & 0.4840 & 0.7923 & 0.2222 & --- \\
\bottomrule
\end{tabular}}
\end{table*}

\paragraph{Diagnostic threshold sweep for external detectors.}
The validation-$F_1$ rule selects near-degenerate thresholds for both external detectors ($\tau = 0.002$ for ProtectAI, $\tau = 0.986$ for DeBERTa-PI), a sign that their score distributions are mismatched to the benchmark rather than that a useful operating point exists. Under these diagnostic settings (Table~\ref{tab:external_results}), neither detector attains high $F_1$ with low benign FPR across axes: tuning ProtectAI down raises IID $F_1$ at the cost of higher hard-benign and structural-shift FPR, while tuning DeBERTa-PI up lowers hard-benign FPR but leaves structural-shift FPR saturated at $1.000$. The default-boundary failures are thus not threshold-placement artifacts; they reflect how these detectors separate benign from adversarial inputs under PIDS-Bench conditions.

\paragraph{Source-origin composition of the IID test split.}
The IID test split contains 2{,}984 real-source and 934 template-derived examples (\texttt{source\_type} field). Because template-derived text is lexically more uniform, aggregate IID $F_1$ overstates real-source performance: for every internal detector, template-only $F_1$ exceeds real-only $F_1$ (DeBERTa-v3-FT $0.9974$ vs.\ $0.9825$; wider gaps for DistilBERT and TF-IDF; Appendix~\ref{app:source_iid}). We therefore report real-source $F_1$ as the primary IID measure throughout. The effect is confined to the IID split; the hard-benign and distribution-shift sets are built independently of the template pool.

\paragraph{Per-domain breakdown under domain shift.}
Aggregate domain-shift $F_1$ hides opposite failure modes: DistilBERT's aggregate $0.8886 \pm 0.0214$ masks the code-domain recall collapse and the legal-domain false-positive inflation analyzed in Section~\ref{sec:distshift} (Table~\ref{tab:domain_ood_breakdown}): one failure admits attacks, the other blocks legitimate traffic, and a single domain-shift $F_1$ conceals both.

\paragraph{Hard-benign score distribution.}
Over-defense reflects high-confidence misclassification, not boundary uncertainty. For DeBERTa-v3-FT on the hard-benign set (seed~42), the injection-probability distribution is strongly bimodal: $62.8\%$ of inputs score $\leq 0.1$ and $32.1\%$ score $\geq 0.9$, with little mass between. The over-blocked inputs sit deep in the injection region, not just above the boundary, so no threshold adjustment recovers them.

\paragraph{Hard-benign over-defense by source origin.}
Table~\ref{tab:hardbenign_origin} decomposes hard-benign FPR into the 872 externally-sourced and 600 curated examples at $\tau = 0.5$. Both subsets show elevated FPR for all three internal detectors, but the pattern varies: DeBERTa-v3-FT is most provenance-dependent ($0.3144 \pm 0.0525$ externally-sourced vs.\ $0.4813 \pm 0.0855$ curated), DistilBERT over-blocks both heavily ($0.4606$ vs.\ $0.4843$), and TF-IDF over-blocks externally-sourced inputs \emph{more} ($0.4622$ vs.\ $0.3317$). Crucially, no detector confines its over-blocking to the curated worst case: all three produce substantial false positives on externally-sourced, topic-filtered prompts from public corpora. We term this asymmetry \emph{provenance-sensitive over-defense}.

\begin{table}[h]
\centering
\caption{Hard-benign false-positive rate by source origin, at $\tau = 0.5$ (five-seed mean $\pm$ std for the neural models; TF-IDF deterministic). The set contains 872 externally-sourced and 600 curated benign examples. Per-source rates are computed by joining each model's hard-benign predictions to the released \texttt{hard\_benign\_test.csv} on \texttt{source\_type}.}

\label{tab:hardbenign_origin}
\small
\setlength{\tabcolsep}{1pt}
\resizebox{\columnwidth}{!}{%
\begin{tabular}{lccc}
\toprule
Model & Aggregate & Externally-sourced & Curated \\
\midrule
TF-IDF + LR   & 0.4090 & 0.4622 & 0.3317 \\
DistilBERT & $0.4702 \pm 0.0823$ & $0.4606 \pm 0.0639$ & $0.4843 \pm 0.0793$ \\
DeBERTa-v3-FT & $0.3825 \pm 0.0474$ & $0.3144 \pm 0.0525$ & $0.4813 \pm 0.0855$ \\
\bottomrule
\end{tabular}%
}
\end{table}

\paragraph{Hard-benign label quality.}
The false-positive rates reported above rest on the blind two-annotator audit of Section~\ref{sec:hardbenign}, which reproduced the benign judgment underlying the measurement.

\paragraph{Per-subtype detection.}
Both fine-tuned detectors recall injections at high and roughly uniform rates across all six subtypes (DeBERTa-v3-FT $0.958$--$1.000$, DistilBERT $0.960$--$1.000$; \texttt{direct\_override} lowest for both), with low benign FPR on this set ($0.013$ and $0.029$; full breakdown in Appendix~\ref{app:subtype}). Attack-subtype identity is therefore not a primary failure axis: the substantial benign false-positive rates elsewhere are not mirrored by any per-subtype detection weakness, which isolates over-defense as a property of benign input provenance rather than attack type.

\subsection{Why High $F_1$ Does Not Imply Low Deployment Risk}
\label{sec:sweep}
The hard-benign and structural-shift results expose a failure that held-out $F_1$ does not capture. Throughout this section, hard-benign FPR refers to the externally-sourced subset ($n = 872$) unless noted; curated and aggregate figures appear only where a worst-case comparison is relevant.

One account of these errors is that the fine-tuned detectors key on lexical and structural features that co-occur with injections in training but also appear in legitimate security documentation, policy text, and structurally unusual benign prompts. Two observations support it: externally-sourced hard-benign false positives carry security-adjacent vocabulary without any override instruction yet receive near-certainty injection scores, and structural-shift benign FPR stays elevated on benign seeds drawn from the hard-benign set, so the surface-form associations persist independently of override content. This matches evidence that neural classifiers can reach strong in-distribution accuracy through superficial heuristics rather than robust semantic features~\cite{mccoy2019right}. We have not tested the account directly through probing or attribution, and present it as a hypothesis consistent with the evidence rather than an established mechanism.

\begin{figure}[h]
\centering
\includegraphics[width=\columnwidth]{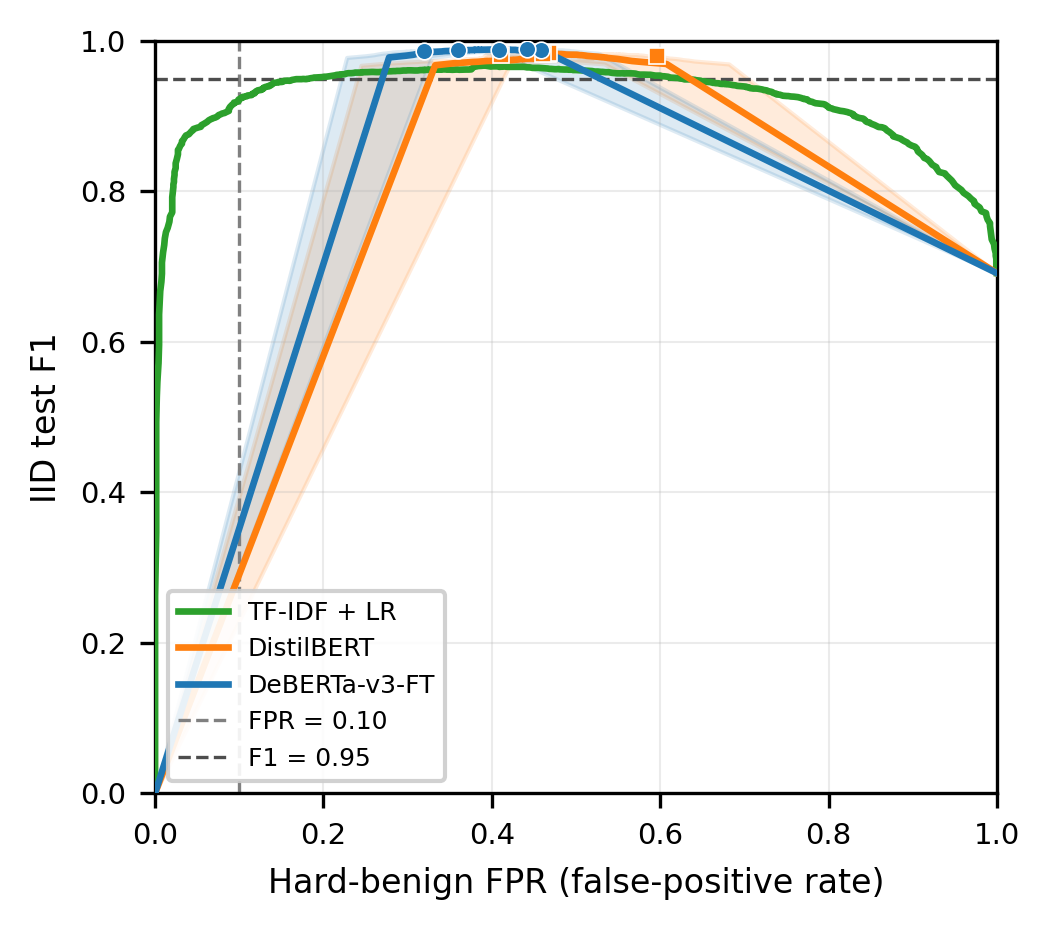}
\caption{Threshold sweep for the three internal learned baselines: each curve traces test $F_1$ against aggregate hard-benign FPR as $\tau$ sweeps $[0,1]$ (five-seed mean per architecture, with a shaded $\pm 1$ standard-deviation band that is narrow throughout, reflecting low cross-seed variance). Markers show the validation-tuned operating points of the two fine-tuned detectors across the five seeds; they cluster tightly for the same reason. The dashed reference lines mark the joint constraint hard-benign FPR $\leq 0.10$ and test $F_1 \geq 0.95$; across the swept grid and all five seeds, no curve enters this region.}
\label{fig:sweep}
\end{figure}

To test whether the hard-benign errors are a thresholding artifact, we swept $\tau$ from $0$ to $1$ for all three internal baselines across five seeds, recording test $F_1$ and hard-benign FPR at each step (Figure~\ref{fig:sweep}). The $F_1 \geq 0.95$ floor is evaluated on the mixed held-out $F_1$ of Table~\ref{tab:internal_results}; on the stricter real-source measure, TF-IDF already falls below it ($0.9467$), so the constraint is if anything lenient for that detector. Subject to it, no model reaches externally-sourced hard-benign $\mathrm{FPR} \leq 0.10$ with $F_1 \geq 0.95$: the lowest attainable externally-sourced FPR is $0.226 \pm 0.058$ for DeBERTa-v3-FT, $0.277 \pm 0.099$ for DistilBERT, and $0.241$ for TF-IDF. The $0.10$ target is itself permissive (a diagnostic reference rather than a deployment requirement, higher than many gateway applications would accept), so a stricter value only widens the gap reported here. The result holds per seed: the per-seed minimum spans $0.150$--$0.311$ for DeBERTa-v3-FT and $0.123$--$0.388$ for DistilBERT, so no single seed of either neural architecture reaches $0.10$. It also holds under the aggregate hard-benign definition (Figure~\ref{fig:sweep}). The detection/over-defense trade-off is therefore not an artifact of threshold placement.

The threshold behavior reinforces this. The validation-tuned threshold of the neural detectors is unstable across seeds (Table~\ref{tab:threshold_rule_comparison}), yet the resulting $(F_1, \mathrm{FPR})$ operating points cluster tightly, because test $F_1$ is nearly flat across the high-scoring region (Figure~\ref{fig:sweep}). Instability in the threshold does not translate into unstable performance, and no single stable threshold separates the two error types. The constrained-rule analysis closes the argument: minimizing externally-sourced FPR subject to the $F_1$ floor still leaves every architecture far above $0.10$ (Table~\ref{tab:threshold_rule_comparison}). The trade-off is a property of the learned decision surface, not of threshold calibration.

\begin{table}[h]
\centering
\caption{Threshold-rule comparison for the three internal learned baselines. The validation-$F_1$ rule is the per-seed threshold-selection rule of Section~\ref{sec:thresholding}; the $\tau$ column gives the range of validation-tuned thresholds across the five seeds (or the single value for the deterministic baseline), and Test $F_1$ and ES hb-FPR are five-seed mean $\pm$ standard deviation. The constrained rule reports the minimum externally-sourced hard-benign FPR attainable subject to test $F_1 \geq 0.95$. No architecture reaches the $0.10$ target under the constrained rule, and the validation-$F_1$ threshold is stable for the linear baseline but highly variable for the fine-tuned neural detectors.}
\label{tab:threshold_rule_comparison}
\scriptsize
\setlength{\tabcolsep}{3pt}
\resizebox{\columnwidth}{!}{%
\begin{tabular}{lcccc}
\toprule
& \multicolumn{3}{c}{Validation-$F_1$ rule} & Constrained ($F_1 \geq 0.95$) \\
\cmidrule(lr){2-4}\cmidrule(lr){5-5}
Model & $\tau$ range & Test $F_1$ & ES hb-FPR & Min ES hb-FPR \\
\midrule
TF-IDF + LR   & $0.518$ & $0.9672$ & $0.4507$ & $0.241$\textsuperscript{a} \\
DistilBERT    & $0.112$--$0.664$ & $0.9831 \pm 0.0017$ & $0.471 \pm 0.062$ & $0.277 \pm 0.099$ \\
DeBERTa-v3-FT & $0.010$--$0.904$ & $0.9882 \pm 0.0008$ & $0.328 \pm 0.064$ & $0.226 \pm 0.058$ \\
\bottomrule
\end{tabular}%
}
\vspace{1mm}
\scriptsize
ES hb-FPR = externally-sourced hard-benign FPR ($n = 872$). \textsuperscript{a}TF-IDF peak $F_1$ ($0.967$) clears the $0.95$ floor only within a narrow threshold band, so its constrained minimum is attained over a small feasible region.
\end{table}

\subsection{Hard-Negative Training Ablation}
\label{sec:ablation}
To test whether over-defense can be reduced through targeted training data, we augment the training set with 419 hard-negative examples (curated benign prompts containing security-adjacent language) and retrain both fine-tuned detectors under identical hyperparameters. The pool is built from training-side seeds only and deduplicated against all frozen evaluation sets (Section~\ref{sec:taxonomy}), ensuring zero overlap with held-out data.

\begin{figure}[h]
\centering
\includegraphics[width=\columnwidth]{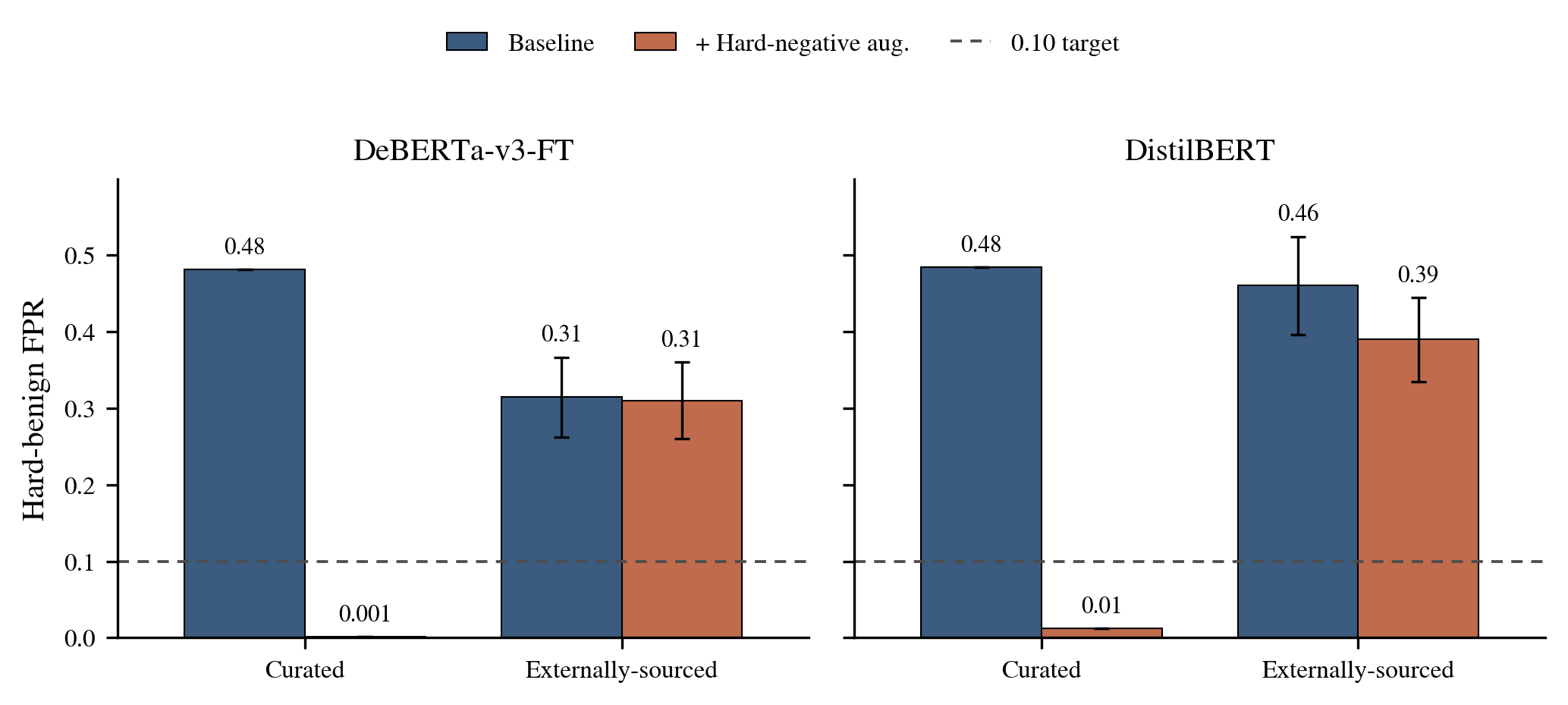}
\caption{Hard-benign false-positive rate by provenance for the two fine-tuned detectors, baseline versus hard-negative augmentation (five-seed means; Table~\ref{tab:ablation}). Error bars are $\pm 1$ standard deviation across seeds on the externally-sourced subset; curated rates have cross-seed variation smaller than the marker. Augmentation collapses curated over-defense to near zero (0.481$\rightarrow$0.001 for DeBERTa-v3-FT, 0.484$\rightarrow$0.012 for DistilBERT) while the externally-sourced rate remains far above the 0.10 target in both architectures (0.314$\rightarrow$0.310, 0.461$\rightarrow$0.390). The asymmetry, mitigation where it is least needed and persistence on the subset closest to real security-adjacent traffic, is \emph{provenance-sensitive over-defense}.}
\label{fig:provenance}
\end{figure}

The central result is an asymmetry between the two hard-benign subsets (Table~\ref{tab:ablation}, Figure~\ref{fig:provenance}). Augmentation nearly eliminates curated over-defense: DeBERTa-v3-FT falls from $0.4813 \pm 0.0855$ to $0.0010 \pm 0.0022$, and DistilBERT from $0.4843 \pm 0.0793$ to $0.0123 \pm 0.0113$. The externally-sourced subset barely moves: DeBERTa-v3-FT goes from $0.3144 \pm 0.0525$ to $0.3101 \pm 0.0502$ (within noise), and DistilBERT from $0.4606 \pm 0.0639$ to $0.3899 \pm 0.0551$ (a modest decrease). The aggregate reductions (DeBERTa-v3-FT $0.3825 \to 0.1841$; DistilBERT $0.4702 \to 0.2360$) are thus driven almost entirely by the curated collapse: the curated examples most resemble the augmentation distribution, while externally-sourced prompts from public corpora remain largely unaffected.

The improvement is training-driven, not a threshold effect. Evaluating each baseline at its augmented counterpart's validation-tuned threshold leaves the aggregate hard-benign FPR near its baseline level, so the reduction follows from the augmented data rather than threshold movement. Augmentation carries a modest domain-shift cost (DeBERTa-v3-FT domain-shift $F_1$ $0.9604 \to 0.9473$; DistilBERT $0.8886 \to 0.8620$) while leaving in-distribution $F_1$ unchanged.

\begin{figure}[h]
\centering
\includegraphics[width=\columnwidth]{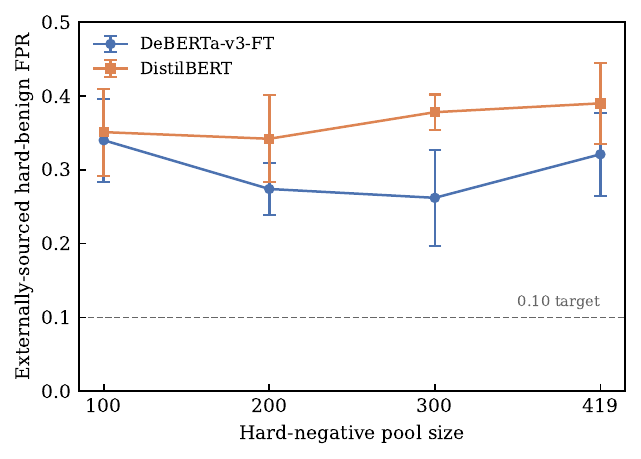}
\caption{Externally-sourced hard-benign false-positive rate across hard-negative pool sizes (five-seed mean $\pm$ standard deviation, $\tau = 0.5$). Enlarging the curated pool produces no systematic reduction in externally-sourced over-defense for either architecture; all points overlap within one standard deviation and remain far above the $0.10$ target. Pool-size runs are trained independently of the main augmentation experiment (Table~\ref{tab:ablation}), so the 419-pool point differs from the corresponding Table~\ref{tab:ablation} value within run-to-run variation.}
\label{fig:poolsweep}
\end{figure}

To rule out pool capacity as the source of this asymmetry, we swept the hard-negative pool size over $\{100, 200, 300, 419\}$ for both architectures, five seeds each, evaluating externally-sourced hard-benign FPR at the fixed threshold. Externally-sourced over-defense shows no systematic dependence on pool size (Figure~\ref{fig:poolsweep}): across $\{100, 200, 300, 419\}$ the per-pool means vary only within cross-seed noise, with all points overlapping within one standard deviation for both architectures, and in neither case does enlarging the pool drive the rate downward toward the $0.10$ target. Enlarging the curated pool does not extend its benefit to externally-sourced prompts. Augmentation drawn to match the externally-sourced distribution is a distinct intervention we do not evaluate; we identify it as future work (Section~\ref{sec:future}) and do not claim hard-negative augmentation is inherently limited, only that curated augmentation does not transfer across provenance at any pool size tested.

\begin{table}[h]
\centering
\caption{Effect of hard-negative augmentation on the two fine-tuned detectors (five-seed mean). Aggregate and externally-sourced hard-benign FPR are reported at $\tau = 0.5$; curated FPR follows the per-source protocol of Table~\ref{tab:hardbenign_origin}. Augmentation nearly eliminates curated over-defense in both architectures while leaving externally-sourced over-defense largely intact.}
\label{tab:ablation}
\small
\setlength{\tabcolsep}{4pt}
\resizebox{\columnwidth}{!}{%
\begin{tabular}{lcccc}
\toprule
& \multicolumn{2}{c}{DeBERTa-v3-FT} & \multicolumn{2}{c}{DistilBERT} \\
\cmidrule(lr){2-3}\cmidrule(lr){4-5}
Metric & Baseline & +HardNeg & Baseline & +HardNeg \\
\midrule
IID $F_1$ & $0.9882$ & $0.9890$ & $0.9831$ & $0.9833$ \\
hb-FPR (aggregate) & $0.3825$ & $0.1841$ & $0.4702$ & $0.2360$ \\
hb-FPR (externally-sourced) & $0.3144$ & $0.3101$ & $0.4606$ & $0.3899$ \\
hb-FPR (curated) & $0.4813$ & $0.0010$ & $0.4843$ & $0.0123$ \\
Domain-shift $F_1$ & $0.9604$ & $0.9473$ & $0.8886$ & $0.8620$ \\
\bottomrule
\end{tabular}%
}
\end{table}

These results show over-defense is partially addressable through targeted augmentation, but the effect concentrates where deployment needs it least: it clears false positives on curated, template-like inputs while leaving externally-sourced over-defense, the behavior closest to real security-adjacent traffic, in place. The asymmetry holds across both fine-tuned architectures, indicating a property of the augmentation strategy rather than of a single model.

\subsection{Disentangling Provenance from Structure}
\label{sec:confound}
Provenance-sensitive over-defense raises a mechanistic question: does the gap reflect source origin itself, or only the instruction-like surface structure that externally-sourced prompts tend to carry? If the latter, the effect would reduce to a structural property. We separate the two by tagging each externally-sourced hard-benign row for instruction-like structure and decomposing the false-positive rate by structure and source corpus jointly.

The tag is deliberately conservative and independent of every evaluated detector, so the measurement does not reduce to one classifier's heuristics. It fires only on explicit instruction constructions from three families: persona assignment (\emph{act as a\ldots}, \emph{pretend to be\ldots}), fictional or hypothetical framing used as a directive (\emph{imagine you are\ldots}), and command-execution phrasing pairing an imperative verb with an execution object (\emph{run the following script}, \emph{invoke the API}). Each requires a constructed imperative, not a bare keyword. Pattern families an audit found to misfire on benign text (encoding cues and instruction-override phrasing among them) are excluded, so the tag does not inherit the false positives it is meant to study.

Structure is a strong and consistent driver. Across the externally-sourced rows, structure-tagged prompts are misclassified far more often than unstructured ones (DeBERTa-v3-FT $\mathrm{FPR} = 0.591$ vs.\ $0.247$; DistilBERT $0.699$ vs.\ $0.413$), and the effect holds within a single corpus, where source is fixed: in the LMSYS plain slice, DeBERTa-v3-FT blocks $0.627$ of structured against $0.383$ of unstructured prompts. It is therefore not an artifact of comparing corpora of differing composition.

Source origin nonetheless adds an effect that structure does not absorb. Partial correlations controlling for the other factor give $0.248$ for structure given source ($p \approx 1\times10^{-13}$) and $0.169$ for source given structure ($p \approx 5\times10^{-7}$); both are significant, with structure the larger. A logistic model agrees: an LMSYS corpus indicator retains a large positive coefficient once structure is included. Over-defense is present across all three corpora rather than concentrated in one, with DeBERTa-v3-FT corpus-level FPR of $0.370$ (LMSYS), $0.183$ (OpenAssistant), and $0.047$ (Dolly). For LMSYS, which contributes no benign training examples, this rate is measured on a source entirely unseen during fine-tuning; for OpenAssistant and Dolly, whose seeds appear in training only as paraphrases, the evaluation rows are disjoint from training with verified zero text overlap. Persistent over-blocking on these topic-filtered rows therefore cannot be explained by memorization of the specific benign examples seen in training. A source-identity probe confirms the corpora are only moderately separable by content: a length-only classifier falls below the majority baseline ($0.549$ against a $0.761$ majority baseline), and a text classifier reaches $0.794$ accuracy but $0.668$ balanced accuracy. The over-defense flag correlates more with structure ($r = 0.284$) than source identity ($r = 0.220$).

The evidence thus supports neither extreme reading. The gap is not purely structural, since source retains a significant effect once structure is controlled; nor is provenance incidental, since corpus-level rates and the retained coefficient reflect a genuine source component. The account is intermediate: instruction-like structure is the larger driver and operates within every provenance, while source origin adds a smaller, statistically independent effect. These estimates are most reliable for the well-powered cells; sparsely populated slices are retained in the released analysis but not interpreted.

\subsection{Operating-Point Analysis}
\label{sec:operating}
The threshold sweep (Section~\ref{sec:sweep}) traces the trade-off across the full operating range; this section complements it at each detector's validation-$F_1$-tuned threshold, the deployment-realistic point, and separates threshold-independent ranking quality from threshold-dependent behavior. Hard-benign FPR here is the aggregate over the full set ($n = 1{,}472$).

\paragraph{Ranking quality is near-ceiling.}
Both fine-tuned detectors rank injections above benign inputs almost perfectly (area under the precision--recall curve, PR-AUC, $0.9985$ for DeBERTa-v3-FT, $0.9979$ for DistilBERT). In the high-specificity region that matters for deployment, the partial ROC-AUC below $\mathrm{FPR} = 0.10$ is $0.850$ for DeBERTa-v3-FT and $0.937$ for DistilBERT: the ordering reverses here, with DistilBERT ranking more cleanly in the low-FPR regime despite trailing elsewhere. Attack recall stays high at strict thresholds: at an IID benign FPR of $0.01$, DeBERTa-v3-FT recalls $0.984$ of injections and DistilBERT $0.963$, both above $0.99$ by an IID FPR of $0.05$. The over-defense reported throughout is thus not a ranking failure but a consequence of where security-adjacent benign inputs fall on the score scale.

\paragraph{Precision collapses at realistic base rates.}
Precision depends on how common injections are, and in deployment they are rare. As the benign-to-injection base rate rises from $90{:}10$ to a realistic $999{:}1$, the precision implied at each detector's operating point drops sharply: DeBERTa-v3-FT from $0.893$ to $0.070$, DistilBERT from $0.813$ to $0.038$ (full breakdown in Appendix~\ref{app:base_rate}). At the base rates of real traffic, most positive predictions are false alarms even for a detector exceeding $F_1 = 0.98$ in-distribution, a cost aggregate $F_1$ hides.

\paragraph{An oracle threshold does not resolve the trade-off.}
To test whether the $0.10$ hard-benign FPR target, or a stricter $0.05$ reference, is attainable in principle, we build a descriptive frontier that selects the threshold on the hard-benign set itself: an oracle requiring label access no deployed detector has, and one that favors the detector. Degenerate thresholds, where the detector abstains almost entirely (attack recall below $0.50$), are excluded, since a low FPR bought by abstention is not an operating point. Even so, no threshold brings DeBERTa-v3-FT's aggregate hard-benign FPR to $0.05$ at usable recall; the lowest rate with recall at least $0.50$ is $0.155$, and the $0.10$ target is met only at the boundary, with maximum test recall $0.922$. DistilBERT admits a feasible threshold for both targets, but at recall $0.789$. Because these thresholds are chosen with hard-benign labels in view, they upper-bound what any leakage-free threshold could reach, placing the constraint in the detectors' score geometry, not in threshold search.

\subsection{Calibration Does Not Explain Over-Defense}
\label{sec:calibration}
The internal detectors are trained independently, and their decision thresholds are not mutually calibrated. This invites an alternative reading of the over-defense reported above: that it reflects overconfident probability estimates rather than the geometry of the learned decision boundary. We test that reading with temperature scaling (a monotonic, validation-only recalibration that rescales probabilities without reordering inputs).

Both fine-tuned detectors are overconfident. The temperature that minimizes validation negative log-likelihood sits well above unity, at $T = 2.02 \pm 0.07$ for DeBERTa-v3-FT and $T = 1.82 \pm 0.06$ for DistilBERT (five-seed mean $\pm$ standard deviation). Scaling roughly halves the expected calibration error (ECE), from $0.01139$ to $0.00515$ for DeBERTa-v3-FT and from $0.01137$ to $0.00487$ for DistilBERT, while the Brier score barely moves (Table~\ref{tab:calibration}). The Brier gap is expected: Brier is dominated by refinement, which temperature scaling cannot alter, and with ROC-AUC near $0.998$ the refinement component is already close to ceiling. The probability estimates are, then, measurably better calibrated after scaling.

The over-defense is not. The aggregate hard-benign false-positive rate is identical before and after scaling: $0.3825$ for DeBERTa-v3-FT and $0.4702$ for DistilBERT under both raw and recalibrated scores. This is expected by construction: temperature scaling divides the logits by a positive constant, leaving their sign unchanged, so the probability-$0.5$ boundary is a fixed point and any threshold set at $0.5$ is untouched.

The substantive point is not that this one threshold survives, but that no monotonic recalibration can relocate the over-blocked inputs at all. As Section~\ref{sec:results} shows, these inputs are scored deep in the injection region ($32.1\%$ above $0.9$ for DeBERTa-v3-FT), and a monotonic transform cannot carry a score that far from the boundary back across it. The raw-scale threshold sweep of Section~\ref{sec:sweep} already showed that no operating point on the original scale escapes the constraint; recalibrating the scale is no more successful. Over-defense therefore lies in where these inputs fall on the learned decision surface, not in the calibration of the score scale.

\begin{table}[h]
\centering
\caption{Calibration of the two fine-tuned detectors before and after temperature scaling (five-seed mean). Temperature is fitted on the validation split only, and ECE uses 15 equal-width confidence bins. Scaling roughly halves ECE but leaves the aggregate hard-benign false-positive rate unchanged (as expected, since the probability-$0.5$ boundary is invariant under temperature scaling), indicating that over-defense is not a calibration artifact.}
\label{tab:calibration}
\setlength{\tabcolsep}{5pt}
\renewcommand{\arraystretch}{1.15}
\resizebox{\columnwidth}{!}{%
\begin{tabular}{@{}lcccc@{}}
\toprule
& \multicolumn{2}{c}{DeBERTa-v3-FT} & \multicolumn{2}{c}{DistilBERT} \\
\cmidrule(lr){2-3}\cmidrule(lr){4-5}
Metric & Raw & Scaled & Raw & Scaled \\
\midrule
Temperature $T$ & \multicolumn{2}{c}{$2.02 \pm 0.07$} & \multicolumn{2}{c}{$1.82 \pm 0.06$} \\
Brier score     & 0.0116 & 0.0108 & 0.0153 & 0.0148 \\
ECE             & 0.01139 & 0.00515 & 0.01137 & 0.00487 \\
Hard-benign FPR & 0.3825 & 0.3825 & 0.4702 & 0.4702 \\
\bottomrule
\end{tabular}%
}
\end{table}

\subsection{Feature-Level Evidence for Over-Defense}
\label{sec:features}
The TF-IDF + logistic-regression baseline permits direct inspection of the features driving its injection decisions. From the seed-42 model we extracted the highest positive-weight features for the injection class and measured their prevalence in the externally-sourced hard-benign subset ($n = 872$).

Setting aside the very highest-weight features, which are generic high-frequency tokens (\textit{and}, \textit{the}, \textit{you}), a clear security cluster emerges, with coefficients between $4.06$ and $5.88$ (Appendix~\ref{app:tfidf}): system-prompt and tool references (\textit{tool}, \textit{internal}, \textit{system}), instruction-override terms (\textit{disregard}, \textit{guidelines}, \textit{directives}, \textit{instructions}), and concealment language (\textit{concealed}, \textit{hidden}). These terms are not confined to curated stress inputs; they recur in genuine externally-sourced benign prompts, where instruction-override terms appear in $198$ examples ($22.7\%$) and system-, tool-, or context-reference terms in $342$ ($39.2\%$). The lexical overlap that drives over-defense is therefore a property of real security-adjacent language, not an artifact of curated templates. Two cases make this concrete: \emph{``Explain why the phrase `disregard your system prompt' is risky in a legal research tool''} receives an injection probability of $0.717$, and \emph{``Help me write a system prompt for a legal research tool that resists role-hijacking attempts''} receives $0.850$.

This evidence exists only for the interpretable linear baseline; DistilBERT and DeBERTa-v3-FT cannot be inspected the same way. It documents a concrete mechanism for the linear model only; the over-defense itself appears in all three internal detectors. That the fine-tuned transformers may rely on analogous associations is consistent with their error pattern, the high-confidence misclassification of security-vocabulary inputs (Section~\ref{sec:results}) and the persistent structural-shift FPR, but is not demonstrated here; direct feature attribution or probing of their internal representations is left to future work.

\subsection{Statistical Analysis}
\label{sec:stats}
The point estimates throughout are five-seed means, and the accompanying standard deviations measure stability across random initialization. This captures one source of uncertainty but not the other: a finite stress set introduces sampling uncertainty that seed variance does not reflect. We therefore report example-level bootstrap confidence intervals ($10{,}000$ resamples) alongside the seed standard deviations, a paired analysis of the augmentation effect, and sampling intervals for the external detectors, which have no seed variation but are still evaluated on finite sets.

\paragraph{Headline confidence intervals.}
The $95\%$ bootstrap intervals for the main metrics (Appendix~\ref{app:cis}) are narrow and agree with the seed standard deviations, confirming that the headline estimates are not sample artifacts: DeBERTa-v3-FT's externally-sourced hard-benign FPR is $0.3144\ [0.2881, 0.3424]$, bounded well away from both the curated rate and the $0.10$ target. We omit example-level intervals for ROC-AUC, where both detectors sit near the $1.0$ ceiling and the bootstrap becomes unreliable against the bound; seed-level variation (Table~\ref{tab:internal_results}) is the appropriate measure there.

\paragraph{The augmentation effect.}
We test the augmentation effect with a paired comparison across the five seeds, combining a paired $t$-test, the Wilcoxon signed-rank test, and an example-level bootstrap (Table~\ref{tab:paired}), so no conclusion rests on a single test's assumptions. Two design features shape the reading. With five paired observations, the Wilcoxon $p$-value cannot fall below $0.0625$; for the large curated and aggregate effects, that floor, not a weak signal, is why the test reports $0.0625$, so we treat the bootstrap as primary. And the two architectures respond differently on the externally-sourced subset, which we report rather than average away.

On the curated subset the effect is large and consistent across all three tests, driving both detectors to near zero. The externally-sourced subset separates them. For DeBERTa-v3-FT the change is null: a paired difference of $-0.004$ with a bootstrap interval straddling zero ($t$-$p = 0.90$). Curated augmentation therefore does not transfer to externally-sourced inputs for this architecture. For DistilBERT the change is a small reduction that does not reach significance across seeds: the paired difference is $-0.071$, but the parametric and rank tests fall short at five seeds ($p = 0.13$ and $0.125$), even as the example-level bootstrap interval excludes zero. Because augmentation is applied per seed, the seed-level tests are the appropriate measure of the effect, so we read the reduction as suggestive rather than established. Neither closes the gap: DistilBERT's externally-sourced FPR after augmentation is still $0.3899$, far above $0.10$, and DeBERTa-v3-FT's is unmoved. The curated--externally-sourced asymmetry thus holds for both architectures; only its magnitude on the externally-sourced side is architecture-dependent.

\paragraph{Sampling uncertainty for external detectors.}
The external and broad-safety detectors run once at fixed operating points, so they carry no seed variance, but their rates remain subject to sampling uncertainty on finite sets (Appendix~\ref{app:ext_cis}). The per-origin split is informative beyond confirming the point estimates: DeBERTa-PI, an injection-trained detector, blocks more externally-sourced than curated benign inputs ($0.825$ vs.\ $0.773$), the same provenance ordering seen in the internal detectors, whereas Llama Guard~3-1B, trained for general safety, is lowest on the curated subset. The provenance-sensitive ordering therefore appears in an off-the-shelf injection detector, not only in the models trained here, which points to injection-specific training rather than safety filtering in general as its source.

\begin{table}[h]
\centering
\caption{Paired analysis of the hard-negative augmentation effect across five seeds. \capDelta{} is augmented minus baseline; a negative value is a reduction in FPR. The bootstrap column is the example-level $95\%$ interval on the difference. The Wilcoxon $p$-value floor at five seeds is $0.0625$. Baseline and augmented values match Table~\ref{tab:ablation}.}
\label{tab:paired}
\renewcommand{\arraystretch}{1.2}
\resizebox{\columnwidth}{!}{%
\begin{tabular}{@{}llccccc@{}}
\toprule
Model & Subset & Base $\to$ Aug & $\Delta$ & $t$-$p$ & Wilc.\ $p$ & Bootstrap CI \\
\midrule
DeBERTa-v3-FT & ext-sourced & $0.314 \to 0.310$ & $-0.004$ & $0.90$   & $1.00$   & $[-0.014, +0.005]$ \\
              & curated     & $0.481 \to 0.001$ & $-0.480$ & $0.0002$ & $0.0625$ & $[-0.515, -0.445]$ \\
              & aggregate   & $0.382 \to 0.184$ & $-0.198$ & $0.0004$ & $0.0625$ & $[-0.218, -0.179]$ \\
\addlinespace[2pt]
DistilBERT    & ext-sourced & $0.461 \to 0.390$ & $-0.071$ & $0.13$   & $0.125$  & $[-0.084, -0.058]$ \\
              & curated     & $0.484 \to 0.012$ & $-0.472$ & $0.0003$ & $0.0625$ & $[-0.504, -0.438]$ \\
              & aggregate   & $0.470 \to 0.236$ & $-0.234$ & $0.0031$ & $0.0625$ & $[-0.253, -0.215]$ \\
\bottomrule
\end{tabular}%
}
\end{table}

\section{Discussion}
\subsection{Positioning Relative to Existing Benchmarks}
\label{sec:positioning}
Benchmarks such as BIPIA and HarmBench measure end-to-end attack success under adversarial prompting: they capture system-level vulnerability but neither isolate the input-level detector nor price its false positives. PIDS-Bench instead treats the detector as a decision boundary and measures attack detection and benign over-blocking jointly, under fixed thresholds, across hard-benign, distribution-shift, and threshold-sweep conditions. Its results are therefore not comparable to end-to-end attack-success figures; they characterize how a gateway detector behaves as a component, an axis those benchmarks leave unmeasured.

\subsection{Failure Profiles and Their Operational Meaning}
\label{sec:profiles}
The central finding is that high in-distribution $F_1$ does not imply low deployment risk: all three internal detectors exceed $0.96$ IID $F_1$ yet flag a substantial share of benign, security-adjacent inputs as injections (Table~\ref{tab:hardbenign_origin}). The three detector groups fail differently.

\paragraph{Fine-tuned injection detectors.}
DeBERTa-v3-FT and DistilBERT pair strong IID performance with high benign false-positive rates under stress, and the failure is not one of threshold placement. No operating point reaches hard-benign FPR $\leq 0.10$ with $F_1 \geq 0.95$ for any architecture or seed (Section~\ref{sec:sweep}), because the over-blocked inputs sit deep in the injection region (roughly a third receive injection probabilities above $0.9$), where no threshold can recover them.

\paragraph{External injection detectors.}
The external detectors over-block heavily at their default boundaries, and tuning relocates the trade-off rather than removing it: ProtectAI's validation-tuned threshold sits near the grid floor and DeBERTa-PI's near the ceiling, and neither yields high $F_1$ with low benign FPR across axes (Table~\ref{tab:external_results}). DeBERTa-PI flags every structural-shift benign input at either threshold. The limitation lies in how these detectors separate benign from adversarial text under PIDS-Bench conditions, not in where the threshold is set.

\paragraph{Broad-safety comparators.}
PromptGuard~2 and Llama Guard~3-1B are general safety classifiers rather than injection detectors. They over-block security-adjacent benign inputs \emph{less} than the injection-trained models (hard-benign FPR $0.15$ and $0.18$ against $0.38$ and $0.47$), but recall far fewer attacks. That the most injection-specialized detectors over-block the most points to over-defense as a consequence of injection-specific training rather than of safety filtering in general.

\subsection{Over-Defense Is a Property of the Decision Surface}
\label{sec:structural}
The over-blocking cannot be tuned away: subject to $F_1 \geq 0.95$, the lowest attainable externally-sourced hard-benign FPR stays well above $0.10$ for every architecture (Table~\ref{tab:threshold_rule_comparison}), and for the strongest detector the residual false positives are high-confidence rather than boundary errors. A consistent account is that the detectors key on lexical and structural cues that co-occur with injections in training but recur in legitimate security discussion: externally-sourced false positives carry security vocabulary without any override instruction yet receive near-certainty scores, and structural-shift benign FPR remains high on security-adjacent seeds that likewise contain no override. This matches evidence that neural classifiers can attain strong in-distribution accuracy through superficial heuristics rather than robust semantic features~\cite{mccoy2019right}; we present it as a hypothesis consistent with the evidence, not a demonstrated mechanism (Section~\ref{sec:limitations}). The hard-negative ablation (Section~\ref{sec:ablation}) reinforces this reading: augmentation nearly removes curated over-defense but leaves the externally-sourced subset largely intact, and the pool-size sweep shows this asymmetry does not close as the pool grows, so the mitigation helps where it is least needed and falls short where deployment needs it most.

\subsection{Distribution-Shift Robustness Is Not a Single Property}
Domain and structural shift surface different failures that aggregate metrics conceal. Under domain shift, DistilBERT's attack recall collapses in the code domain while its benign FPR stays low, whereas TF-IDF holds code recall but inflates benign false positives in the legal domain (Table~\ref{tab:domain_ood_breakdown}): opposite failures with opposite costs, one admitting attacks and the other blocking legitimate traffic, both hidden by a single domain-shift $F_1$. Structural shift is dominated by benign-side inflation, largest under instruction-prefix mimicry: for DeBERTa-v3-FT it drives benign FPR to $0.918 \pm 0.071$, above prompt dilution ($0.796 \pm 0.095$) and JSON wrapping ($0.719 \pm 0.054$), and it is the largest transform in every seed, consistent with its closer mimicry of an authoritative instruction prefix.

\subsection{Implications for Deployment}
\label{sec:implications}
These patterns expose a risk that standard held-out $F_1$ cannot see. In settings that routinely handle security-related requests (enterprise retrieval assistants, developer-support tools, security-education chatbots), a detector tuned for aggregate IID performance may block legitimate discussion at high rates. Four practices follow. First, report hard-benign FPR alongside IID $F_1$: the two must be read together, with the externally-sourced subset the more conservative deployment reference and the aggregate a worst-case characterization rather than an operational rate. Second, include security-adjacent benign inputs in validation, since standard splits do not surface over-defense and cannot inform threshold selection against it. Third, match the detector to the operating condition, weighing the expected input distribution and the relative cost of missed attacks against false positives (Tables~\ref{tab:internal_results} and~\ref{tab:external_results}). Fourth, disaggregate distribution-shift performance along both error axes, since recall collapse and FPR inflation are distinct failures with distinct costs.

\subsection{Future Directions}
\label{sec:future}
Several questions remain. Whether augmentation drawn to match the externally-sourced distribution (rather than the curated pool tested here) can close the residual gap is untested; the pool-size sweep establishes only that scaling a curated pool does not. So is controlled leave-one-source-out retraining: the source-controlled analysis of Section~\ref{sec:confound} is observational, and a source-holdout design remains the direct test of the residual source component. Whether detectors generalize to obfuscation strategies absent from training, such as semantic paraphrase or multi-step encoding, is likewise open. A mechanistic question also remains: whether the fine-tuned transformers rely on the lexical associations identified for the linear baseline, which would require probing or attribution methods suited to their representations. Thresholding conditioned on input provenance or on structural markers such as instruction prefixes or JSON wrappers is also untested: per-class operating points could trade recall against false positives where the single global rule of Section~\ref{sec:thresholding} cannot, though a threshold keyed to attacker-controllable structure may open an evasion channel of its own. The calibration analysis is similarly confined to temperature scaling; isotonic regression and Bayesian Brier calibration remain to be compared against it on both calibration quality and hard-benign FPR, with the caveat that monotone recalibration leaves the attainable operating frontier unchanged (Section~\ref{sec:calibration}), so any movement on over-defense would have to come from non-monotone methods. Finally, extension to multilingual and multi-turn inputs would address deployment-relevant distribution shifts that the present single-turn, English-only protocol does not cover.

\section{Limitations}
\label{sec:limitations}
We set out the boundaries of PIDS-Bench and of our findings plainly, so that each reported figure is read within the scope under which it was measured.

\paragraph{Scope of inputs.}
The benchmark is limited to English, single-turn prompts. Multilingual and multi-turn inputs, both common in deployment, are absent, and the distribution shifts they introduce fall outside what we measure. The evaluation methodology carries over to those settings; the specific numbers do not. We treat both as priority extensions.

\paragraph{Two-annotator hard-benign audit.}
The blind two-annotator audit (Section~\ref{sec:hardbenign}) established the benign labels underlying the false-positive measurement; the finer security-adjacent designation showed only moderate agreement, consistent with these prompts occupying a genuinely ambiguous region. We note that the $\kappa = 0.758$ reported for the six-way subtype audit is a separate measurement (a different rubric on a different sample) and should not be conflated with the hard-benign agreement. The remaining caveat is that both annotators are project-internal; an external audit is left to future work.

\paragraph{Template-derived content in the main split.}
For the two subtypes with the fewest externally-sourced examples, tool injection and encoded attacks, we paraphrase real seeds with a single model (GPT-4o-mini). Such text is systematically easier for the internal detectors: DeBERTa-v3-FT scores $0.9974$ IID $F_1$ on template-derived inputs against $0.9825$ on real-source ones (Appendix~\ref{app:source_iid}). This inflates aggregate IID $F_1$, which is why we report real-source $F_1$ as the primary in-distribution measure. The effect is contained, since template-derived rows are concentrated in training and make up $23.8\%$ of the held-out test split, but PIDS-Bench IID figures should still be compared cautiously against benchmarks built with other generation procedures. The same uniformity bears on the per-subtype recall of Table~\ref{tab:balanced_subtype}. Because a single generator produced the tool-injection and encoded-attack templates, held-out examples of these subtypes share lexical and structural regularities with their training counterparts; seed-family separation prevents variant leakage across partitions but does not remove this shared style. Recall on these two subtypes is therefore best read as an upper bound, and attacks written by other generators or by hand could lower it.

\paragraph{The proposed mitigation is partial.}
Hard-negative augmentation reduces over-defense where deployment needs it least. It nearly clears the curated subset (DeBERTa-v3-FT $0.4813 \to 0.0010$; DistilBERT $0.4843 \to 0.0123$) while barely touching the externally-sourced subset, the case closest to real traffic ($0.3144 \to 0.3101$; $0.4606 \to 0.3899$). Varying the pool size over $\{100, 200, 300, 419\}$ left externally-sourced over-defense unchanged in both architectures within seed variation (Section~\ref{sec:ablation}), so the asymmetry does not stem from pool capacity. What we have not tested is augmentation drawn to match the externally-sourced distribution rather than the curated pool; whether that would close the gap is open (Section~\ref{sec:future}). We therefore treat augmentation as a diagnostic probe of the over-defense mechanism, not a solved mitigation.

\paragraph{Threshold instability.}
The validation-tuned threshold of the fine-tuned detectors swings widely across seeds (Section~\ref{sec:sweep}, Table~\ref{tab:threshold_rule_comparison}), unlike the linear baseline's stable $\tau = 0.518$. Because no single operating point separates the two error types consistently across initializations, any single-threshold result for the neural detectors should be read as one point on an unstable surface rather than a fixed property, itself a deployment concern.

\paragraph{Feature evidence is confined to the linear model.}
The interpretable feature analysis (Section~\ref{sec:features}) covers only TF-IDF with logistic regression. The transformers' error patterns are consistent with their relying on similar lexical and structural cues, but we do not show this directly; probing their internal representations is left to future work. The representation account is therefore a hypothesis the evidence supports, not an established mechanism.

\paragraph{Constructed shift and obfuscation sets.}
The distribution-shift and obfuscation evaluations use constructed rather than naturally sampled traffic. The structural transforms (JSON wrapping, prompt dilution, and instruction-prefix mimicry) are programmatic reformatting operations whose prevalence in real traffic is unknown, and their semantic preservation was checked during construction but not independently annotated, so instruction-prefix mimicry in particular carries that caveat. The obfuscation set applies leetspeak, homoglyph, and zero-width transforms of kinds already present in training, so it measures sensitivity to known perturbations rather than robustness to novel evasion. For domain shift, the injections are GPT-4o-mini generations in domain vocabulary and the benign side comes from external domain corpora; the results thus describe behavior on these constructed and externally-sourced inputs, not on naturally occurring domain traffic.

\paragraph{Hard-benign representativeness.}
The curated portion of the hard-benign set (600 of 1{,}472 rows) was written to concentrate inputs that stress over-defensive detectors. This sharpens diagnostic sensitivity but means the curated subset does not mirror the distribution of security-adjacent language in any specific deployment. Tying hard-benign FPR to an operational false-positive cost would require either real deployment traffic with verified benign labels or an estimate of how often security-adjacent queries arise in the target application, neither of which we provide. The externally-sourced subset (872 rows) and its per-source rates (Table~\ref{tab:hardbenign_origin}) give a more conservative reference; aggregate hard-benign figures should be read as worst-case characterizations, not as predictions of operational false-positive rates.

\section{Conclusion}
PIDS-Bench evaluates prompt-injection detectors on the behavior that standard held-out metrics leave unmeasured: what a detector does to benign, security-adjacent traffic and how it holds up under distribution shift. The benchmark is frozen and checksum-verified. It contains 40{,}479 unique labeled prompts: a 35{,}000-row main corpus and five stress sets, each targeting a distinct failure mode. Counting the 2{,}693 transformed test prompts, the evaluation spans 43{,}172 instances (Section~\ref{sec:composition}). Its central finding is simple to state and hard to see under aggregate $F_1$: a detector can be near-perfect in-distribution and still over-block a large share of legitimate security-related queries.

DeBERTa-v3-FT makes this concrete. It reaches IID $F_1 = 0.9882$ yet, at $\tau = 0.5$, flags $31.4\%$ of externally-sourced security-adjacent benign inputs as injections (Table~\ref{tab:hardbenign_origin}). The behavior is neither a thresholding nor a calibration artifact: across the full sweep and all five seeds, no operating point on the evaluated benchmark meets externally-sourced hard-benign $\mathrm{FPR} \leq 0.10$ and $F_1 \geq 0.95$ together, and temperature scaling sharpens calibration without shifting the false-positive rate. The over-blocked inputs sit near the top of the score distribution, not at its boundary, so no threshold recovers them.

Targeted augmentation helps, but not where it is needed. Hard negatives nearly erase over-defense on the curated subset in both architectures while leaving the externally-sourced subset, the case closest to real traffic, essentially untouched, at a modest cost under domain shift. Enlarging the augmentation pool changes nothing: externally-sourced over-defense stays flat across pool sizes. The one avenue we have not tested is augmentation matched to the externally-sourced distribution rather than to curated worst-case inputs; whether that closes the gap is open. What we can state firmly is narrower: threshold tuning and curated-style augmentation alone do not remove security-adjacent over-blocking.

We call this failure mode \emph{provenance-sensitive over-defense}: over-defense is present on both curated and externally-sourced benign inputs at baseline, but hard-negative augmentation removes it only on the curated side, leaving the externally-sourced side, the case closest to real traffic, largely in place across both fine-tuned architectures, all seeds, and every pool size tested. Domain shift reveals a mirror-image failure on the attack side: DistilBERT's code-domain recall falls to $0.435$ while its benign false-positive rate stays low. Robustness under shift is therefore two-sided, and a single aggregate score hides which side is failing.

The broader lesson for the field is that evaluating a prompt-injection detector means measuring its benign-side behavior on security-adjacent inputs directly, alongside obfuscation robustness and performance under domain and structural shift, and reading stress-set FPR as a characterization of sensitivity, not a forecast of deployment rates. Our results are scoped to English, single-turn inputs; multilingual and multi-turn evaluation remains open. By releasing frozen splits, checksums, and a shared multi-axis protocol, PIDS-Bench gives that evaluation a reproducible foundation to build on.

\appendix
\section{Supplementary Tables}
This appendix collects the confirmatory tables referenced in Section~\ref{sec:results}. Each supports a conclusion stated in the main text; the full breakdowns are provided here for completeness and reproducibility.

\subsection{IID Performance by Source Origin}
\label{app:source_iid}
Table~\ref{tab:source_iid} decomposes IID test $F_1$ into real-source and template-derived subsets for all three internal detectors, confirming that template-derived examples yield uniformly higher $F_1$ and that aggregate IID figures therefore overstate real-source performance (Section~\ref{sec:results}).

\begin{table}[h]
\centering
\caption{IID test performance by source origin (five-seed mean for the neural models; TF-IDF is deterministic). Mixed corresponds to the full 3{,}918-example test split; real-source ($n=2{,}984$) and template-derived ($n=934$) subsets are defined by the \texttt{source\_type} field. All values at $\tau = 0.5$.}
\label{tab:source_iid}
\small
\setlength{\tabcolsep}{2pt}
\begin{tabular}{lccc}
\toprule
Model & Mixed F1 & Real-source F1 & Template F1 \\
\midrule
TF-IDF + LR   & 0.9656 & 0.9467 & 0.9962 \\
DistilBERT    & 0.9831 & $0.9738 \pm 0.0026$ & $0.9980 \pm 0.0007$ \\
DeBERTa-v3-FT & 0.9882 & $0.9825 \pm 0.0009$ & $0.9974 \pm 0.0009$ \\
\bottomrule
\end{tabular}
\end{table}

\subsection{Per-Subtype Attack Recall}
\label{app:subtype}
Table~\ref{tab:balanced_subtype} reports attack recall across the six injection subtypes on the balanced-subtype stress set, confirming that detection is high and roughly uniform across strategies and that attack-subtype identity is not a primary failure axis (Section~\ref{sec:results}).

\begin{table}[h]
\centering
\caption{Per-subtype attack recall on the balanced-subtype stress set (2{,}297 rows: 1{,}097 injection across six subtypes, 1{,}200 benign), five-seed mean $\pm$ standard deviation at each model's validation-tuned threshold.}
\label{tab:balanced_subtype}
\small
\setlength{\tabcolsep}{4pt}
\resizebox{\columnwidth}{!}{%
\begin{tabular}{lcc}
\toprule
Subtype ($n$) & DeBERTa-v3-FT & DistilBERT \\
\midrule
\texttt{direct\_override} (200)        & $0.958 \pm 0.011$ & $0.960 \pm 0.009$ \\
\texttt{contextual\_manip.} (200)      & $0.998 \pm 0.003$ & $0.995 \pm 0.005$ \\
\texttt{roleplay\_attack} (200)        & $1.000 \pm 0.000$ & $1.000 \pm 0.000$ \\
\texttt{data\_exfiltration} (200)      & $0.997 \pm 0.005$ & $1.000 \pm 0.000$ \\
\texttt{tool\_injection} (152)         & $0.988 \pm 0.009$ & $0.984 \pm 0.004$ \\
\texttt{encoded\_attack} (145)         & $0.999 \pm 0.003$ & $1.000 \pm 0.000$ \\
\midrule
Overall balanced $F_1$                 & $0.988 \pm 0.001$ & $0.979 \pm 0.002$ \\
Benign FPR                             & $0.013 \pm 0.002$ & $0.029 \pm 0.005$ \\
\bottomrule
\end{tabular}%
}
\end{table}

\subsection{High-Weight Injection-Class Features}
\label{app:tfidf}
Table~\ref{tab:tfidf_features} lists the highest-weight security- and instruction-related features in the seed-42 TF-IDF + logistic-regression model, documenting the lexical basis of the linear baseline's over-defense discussed in Section~\ref{sec:features}.

\begin{table}[h]
\centering
\caption{Representative high-weight injection-class features in the seed-42 TF-IDF + logistic-regression model, restricted to interpretable security- and instruction-related terms.}
\label{tab:tfidf_features}
\small
\begin{tabular}{lc}
\toprule
Feature & Coefficient \\
\midrule
tool & 5.883 \\
internal & 5.825 \\
disregard & 5.176 \\
system & 5.113 \\
guidelines & 5.033 \\
directives & 4.860 \\
instructions & 4.528 \\
context & 4.527 \\
concealed & 4.486 \\
hidden & 4.055 \\
\bottomrule
\end{tabular}
\end{table}

\subsection{Precision Under Varying Base Rates}
\label{app:base_rate}
Table~\ref{tab:base_rate} gives the precision implied at each fine-tuned detector's validation-tuned operating point as the benign-to-injection base rate varies, quantifying the precision collapse under realistic class imbalance discussed in Section~\ref{sec:operating}.
\begin{table}[h]
\centering
\caption{Precision at each fine-tuned detector's validation-tuned operating point as the benign-to-injection base rate varies, from $90{:}10$ to a very rare $999{:}1$ injection rate.}
\label{tab:base_rate}
\small
\setlength{\tabcolsep}{6pt}
\begin{tabular}{lccc}
\toprule
Base rate (benign:injection) & $90{:}10$ & $99{:}1$ & $999{:}1$ \\
\midrule
DeBERTa-v3-FT & 0.893 & 0.431 & 0.070 \\
DistilBERT    & 0.813 & 0.284 & 0.038 \\
\bottomrule
\end{tabular}
\end{table}

\subsection{Bootstrap Confidence Intervals: Internal Detectors}
\label{app:cis}
Table~\ref{tab:cis} reports example-level bootstrap $95\%$ confidence intervals for the internal detectors, confirming that the headline estimates are stable under resampling and agree with the seed-level variation (Section~\ref{sec:stats}).

\begin{table}[h]
\centering
\caption{Example-level bootstrap $95\%$ confidence intervals ($10{,}000$ resamples) for the internal detectors. Point estimates agree with Tables~\ref{tab:internal_results} and~\ref{tab:hardbenign_origin}. 
ROC-AUC is omitted: near the $1.0$ ceiling the bootstrap interval is unreliable, and seed-level variation (Table~\ref{tab:internal_results}) is reported instead.}
\label{tab:cis}
\small
\setlength{\tabcolsep}{4pt}
\renewcommand{\arraystretch}{1.15}
\resizebox{\columnwidth}{!}{%
\begin{tabular}{@{}llc@{}}
\toprule
Model & Metric & Value $[95\%$ CI$]$ \\
\midrule
DeBERTa-v3-FT & IID $F_1$            & $0.9882\ [0.9859, 0.9922]$ \\
              & Real-source $F_1$    & $0.9825\ [0.9787, 0.9886]$ \\
              & IID benign FPR       & $0.0124\ [0.0070, 0.0162]$ \\
              & hb-FPR (ext-sourced) & $0.3144\ [0.2881, 0.3424]$ \\
              & hb-FPR (curated)     & $0.4813\ [0.4467, 0.5153]$ \\
              & hb-FPR (aggregate)   & $0.3825\ [0.3603, 0.4043]$ \\
\addlinespace[2pt]
DistilBERT    & IID $F_1$            & $0.9831\ [0.9810, 0.9885]$ \\
              & Real-source $F_1$    & $0.9738\ [0.9710, 0.9827]$ \\
              & IID benign FPR       & $0.0245\ [0.0152, 0.0287]$ \\
              & hb-FPR (ext-sourced) & $0.4606\ [0.4317, 0.4899]$ \\
              & hb-FPR (curated)     & $0.4843\ [0.4507, 0.5187]$ \\
              & hb-FPR (aggregate)   & $0.4702\ [0.4482, 0.4920]$ \\
\bottomrule
\end{tabular}%
}
\end{table}

\subsection{Bootstrap Confidence Intervals: External Detectors}
\label{app:ext_cis}
Table~\ref{tab:ext_cis} reports example-level bootstrap $95\%$ confidence intervals for the external and broad-safety detectors at their default operating points, including the per-origin split that shows DeBERTa-PI blocking more externally-sourced than curated benign inputs, the same provenance ordering seen in the internal detectors (Section~\ref{sec:stats}).

\begin{table}[h]
\centering
\caption{Example-level bootstrap $95\%$ confidence intervals for the external and broad-safety detectors at their default operating points. Aggregate values match Table~\ref{tab:external_results}.}
\label{tab:ext_cis}
\renewcommand{\arraystretch}{1.15}
\resizebox{\columnwidth}{!}{%
\begin{tabular}{@{}lccc@{}}
\toprule
Detector & Aggregate & Ext-sourced & Curated \\
\midrule
ProtectAI ($\tau{=}0.5$)  & $0.216\ [0.195, 0.238]$ & $0.083\ [0.065, 0.101]$ & $0.410\ [0.372, 0.448]$ \\
DeBERTa-PI ($\tau{=}0.5$) & $0.804\ [0.783, 0.823]$ & $0.825\ [0.799, 0.849]$ & $0.773\ [0.740, 0.807]$ \\
PromptGuard~2 (86M)       & $0.151\ [0.134, 0.171]$ & $0.102\ [0.083, 0.123]$ & $0.223\ [0.190, 0.258]$ \\
Llama Guard~3-1B          & $0.180\ [0.161, 0.200]$ & $0.237\ [0.209, 0.267]$ & $0.097\ [0.073, 0.120]$ \\
\bottomrule
\end{tabular}%
}
\end{table}

\section*{Data Availability}
The frozen benchmark is publicly available at the project repository,\footnote{\url{https://github.com/ShirePyDev/Prompt-Injection-Detection-System}} released as an immutable tagged version (\texttt{v1.0-pids-bench}, commit \texttt{87dc835}) so that every reported result ties to a verifiable artifact. The release contains the split files, the five stress sets, the hard-negative augmentation pool, the two-annotator audit labels, and the evaluation and analysis code, with SHA-256 checksums recorded in \texttt{FREEZE\_MANIFEST.txt}. Per-source licenses, provenance, and terms are documented in \texttt{DATA\_LICENSES.md}. Because two constituent corpora carry non-commercial licenses, the benchmark is intended for research and non-commercial use.

One subset of the externally-sourced hard-benign rows is drawn verbatim from LMSYS-Chat-1M, whose license does not permit redistribution of the conversation text. These rows are released with the text removed and a SHA-256 fingerprint retained in its place; \texttt{rebuild\_restricted.py} restores them from a licensed copy of LMSYS-Chat-1M by matching each row to its text by fingerprint, which we verified recovers all 664 affected rows exactly. Every other row is released in full, including the Stanford Alpaca and OpenAssistant training rows, which are GPT-4o-mini paraphrases of their seeds rather than verbatim source text and are attributed accordingly.

\section*{Ethical Considerations}
PIDS-Bench contains adversarial prompts, jailbreak attempts, and instruction-injection attacks, some carrying offensive or harmful language. We retain this material deliberately: a prompt-injection detector must be evaluated on the inputs it will actually meet, and filtering out offensive content would make the benchmark unrepresentative of deployment. The content does not reflect the authors' views and is released solely to support research on detecting and mitigating prompt injection. A content note accompanies the artifact, describing its nature and scope so that researchers can handle it appropriately, particularly where examples are displayed or annotated. The annotation described in Section~\ref{sec:benchmark} was confined to label verification by the authors and graduate annotators on already-collected public and constructed text; it involved no human subjects and no personal data.

\section*{Acknowledgment}
The authors thank the graduate annotators who contributed to the attack-subtype annotation audit, and the second annotator who independently re-labeled the hard-benign audit sample.

This work was supported by the National Research Foundation of Korea (NRF) grant (RS-2026-25522067) and by the Institute of Information \& Communications Technology Planning \& Evaluation (IITP) grants (IITP-2024-RS-2024-00397085; Leading Generative AI Human Resources Development, and IITP-2024-RS-2024-00417958; Global Research Support Program in the Digital Field), funded by the Korea government (MSIT).

\end{document}